%% file: main.tex
\documentclass[journal]{IEEEtran}

\usepackage[T1]{fontenc}
\usepackage[utf8]{inputenc}
\usepackage[english]{babel}
\usepackage[babel]{csquotes}
\usepackage{amssymb,amsthm} 
\usepackage{amsfonts,amsmath}
\usepackage{graphicx}
\usepackage[ruled]{algorithm2e} 
\usepackage{cite}
\usepackage{url}
\usepackage{xcolor}
\usepackage[acronym,shortcuts]{glossaries}
\usepackage{float}
\usepackage{hyperref} 
\usepackage{booktabs} 

\input{acro_list} 

\begin{document}

\title{Continuous-Time Analysis of \ac{AFDM}: \\ Pulse-Shaping, Fundamental Bounds and\\ Impact of Hardware Impairments}
\author{Michele~Mirabella$^\ddagger$,~\IEEEmembership{Member,~IEEE}, \,\!
Hyeon Seok Rou$^{\dagger}$,~\IEEEmembership{Member,~IEEE}, \,\!
Pasquale~Di~Viesti$^\ddagger$,~\IEEEmembership{Member,~IEEE},\\
Giuseppe Thadeu Freitas de Abreu$^{\dagger}$,~\IEEEmembership{Senior~Member,~IEEE}, \,\!
and Giorgio~M.~Vitetta$^\ddagger$,~\IEEEmembership{Senior~Member,~IEEE}.
\thanks{
Authors are with: $^\ddagger$University of Modena and Reggio Emilia, Dept. of
Engineering ``Enzo Ferrari'', Via P. Vivarelli 10/1, 41125 Modena (Italy), \ac{CNIT}, and with $^{\dagger}$School of Computer Science and Engineering, Constructor University, 28759 Bremen, Germany. Emails: michele.mirabella@unimore.it, hrou@constructor.university, pasquale.diviesti@unimore.it, gabreu@constructor.university, giorgio.vitetta@unimore.it. Part of this work has been supported by the European \ac{PNRR}. Specifically the project has involved the MOST – Sustainable Mobility National Research Center.} \vspace{-3ex}}
\maketitle

\glsresetall

\begin{abstract}
\Ac{AFDM} has recently emerged as a resilient waveform candidate for high-mobility next-generation wireless systems.
However, current literature mostly focuses on \ac{DT} models, often overlooking effects and hardware non-idealities of actual \ac{CT} signal generation.
In this paper, we bridge this gap by developing a \ac{CT}-analytical framework based on the \ac{AFS} representation, which allows us to demonstrate that strictly bandlimited pulses and subcarrier suppression strategies are essential to maintain the multicarrier structure of the transmitted signal.
In addition, we derive the analytical power spectral density of \ac{AFDM} and compare its spectral characteristics in comparison with those of other multicarrier schemes, taking into account the impact of realistic truncated pulse-shaping.
Furthermore, we analyze the sensitivity of the \ac{CT} model to phase noise, carrier frequency offset, and sampling jitter, providing a theoretical analysis of communication performance.
Finally, we derive closed-form Cram{\'e}r-Rao bounds for channel parameter estimation, showing that the chirped modulation peculiar of \ac{AFDM} increases estimation variance but enables the resolution of Doppler ambiguities.
Our findings lay the necessary theoretical and practical foundations for the implementation of \ac{AFDM} in realistic wireless transceivers.
\end{abstract}

\markboth{IEEE Journal on Selected Areas in Communications}{Submitted paper}

\glsresetall


\setcounter{page}{1}

\glsresetall

\section{Introduction} \label{Section_Introduction}

The standardization of the \ac{6G} of public wireless networks is driven by the need to support extreme mobility scenarios, including high-speed railway systems, \ac{CAV}, \ac{LEO} satellite constellations, \ac{UAVs}, and more.
In such environments, \ac{TF} dispersion results in a \ac{DS}, or doubly-dispersive, fading profile, characterized by high Doppler shifts and multipath delays \cite{wang_6g_2020,Bliss_Govindasamy_2013}, which have been recently investigated for a variety of \ac{6G} scenarios \cite{Rayan_SIM, Rayan_FIM}.

\Ac{OFDM}, the cornerstone of \ac{4G} and \ac{5G} systems, is remarkably robust to frequency-selective fading, but suffers significant performance degradation in high-mobility scenarios.
In particular, \ac{DS} channels result in loss of orthogonality among subcarriers, leading to \ac{ICI} which complicates channel estimation and detection, significantly deteriorating the performance of \ac{OFDM} \cite{wang_performance_2006}.
To address these challenges, novel waveform designs have been proposed, such as the \ac{OTFS} modulation, operating in the invariant \ac{DD} domain and exhibiting excellent robustness to Doppler effects \cite{mohammed_otfs_2023, mirabella_use_2024}.

Recently, however, \ac{AFDM} has emerged as a flexible and promising alternative to \ac{OTFS} \cite{bemani_affine_2023,li_affine_2026,rou2026afdmCSM}, showing lower detection complexity and greater compatibility with \ac{OFDM} \cite{Rou_otfsafdm24,rou2026afdm6g}.
Based on the \ac{DAFT}, \ac{AFDM} employs chirp-periodic basis functions parameterized by a tunable chirp index.
By optimally adjusting this chirp parameter, \ac{AFDM} can align the signal representation with the Doppler profile of the channel, thus achieving full diversity and ensuring that multipath components remain separable in the \ac{AFD} \cite{tao_affine_2025}.
This property makes \ac{AFDM} particularly attractive not only for communications but also for \ac{ISAC}, as it efficiently allows for the estimation of delay and Doppler parameters \cite{luo_novel_2025}.

Despite the growing body of literature on \ac{AFDM}, the majority of existing studies rely exclusively on \ac{DT} matrix-algebraic signal models.
While sufficient for designing digital precoders or detectors \cite{Ranasinghe_JCDEAFDM24}, \ac{DT} models fail to capture \ac{PHY} issues which may impact practical deployment, since, in real hardware, signals are generated and received in \ac{CT}.
It is worth remarking that \ac{CT} signal modeling has already shown its usefulness in the analysis of several digital modulation schemes available in the literature.
In particular, exhaustive \ac{CT} analyses have also been conducted for \ac{OFDM} systems, where such an approach has led to a rigorous characterization of \acp{HWI}, including pulse-shaping effects, \ac{ISI} and \ac{ICI}, synchronization errors, and \ac{PN} \cite{vitetta_wireless_2013, mengali_synchronization_2013}.
Other recent examples include the \ac{FTN} variant of \ac{OFDM} known as \ac{SEFDM} \cite{SarwarTWC2023}, the precursor of \ac{AFDM}, namely, \ac{OCDM} \cite{CuozzoOJCS2024}, frequency-modulated \ac{OFDM} \cite{DiGennaroOJCS2025} and, of course, the more recent \ac{OTFS} modulation.

Beyond interference and synchronization aspects, \ac{CT} modeling has also proven to be a useful tool for the analysis of the peak behavior of \ac{OFDM} signals. In this context, \cite{wulich_comments_2000} exploited a \ac{CT} analytical framework to study the peak factor of multicarrier waveforms, highlighting limitations of \ac{DT} representations. Moreover, \cite{sharif_peak_2003} derived analytical bounds on the \ac{PMEPR} as a function of the oversampling rate and proposed some methods for peak power estimation together with associated error bounds.

In the context of \ac{FTN} multicarrier schemes, a \ac{SEFDM} variant that closely follows the \ac{OFDM} transmission structure, employing a \ac{CP}, \ac{RRC} pulse-shaping at the transmitter and the introduction of virtual carriers was recently proposed in \cite{mirabella_soft_2025,mirabella_joint_2026}. This design allows for a simple \ac{CT} signal representation and significantly simplifies receiver architecture, especially in terms of channel estimation and subsequent detection, while preserving the spectral efficiency gains of \ac{SEFDM} over \ac{OFDM}.

For \ac{OTFS}, recent works such as \cite{mohammed_otfs_2023,mirabella_use_2024} have proposed \ac{CT} frameworks that explicitly account for pulse-shaping and its impact on pilot placement and channel estimation performance. While \cite{mohammed_otfs_2023} focuses on Zak-\ac{OTFS} signaling based on \ac{TF} localized waveforms (i.e., the so-called ``pulsones''), \cite{mirabella_use_2024} adopts a multicarrier-based approach inspired by \ac{OFDM}, employing a \ac{DCP} structure and \ac{RRC} pulses to guarantee \ac{ISI}-free reception in both time and frequency domains.

Similarly, for \ac{OCDM}, the \ac{CT} analysis presented in \cite{omar_performance_2021} has allowed the authors to highlight several non-ideal effects, including \ac{TBI}, \ac{NBI}, as well as to derive analytical expressions for the \ac{PAPR}.

Motivated by the above discussion, this paper develops a \ac{CT} signal model for the \ac{AFDM} modulation, generated by following the approach adopted for multicarrier schemes. The framework highlights aspects which are often overlooked by \ac{DT} formulations, including pulse-shaping, receiver filtering, sampling, \ac{LO} impairments, and fractional delay and Doppler shifts, and their impact on \ac{AFDM} orthogonality, performance bounds for communication, and channel parameter estimation. Furthermore, the \ac{CT} model allows us to analyze implementation-related aspects, such as spectral occupancy, synchronization sensitivity, and the interplay between \ac{PN}, \ac{CFO}, and channel time variations, thereby filling a relevant modeling gap in the \ac{AFDM} literature, and providing tools for realistic performance assessment and advanced transceiver design.

Our main contributions can be summarized as follows:

1) We provide, for the first time, a rigorous derivation of the \ac{AFDM} waveform using a \ac{CT} model based on the \ac{AFS}. This framework is closer to physical reality than \ac{DT} models available in the current \ac{AFDM} literature, and provides insights for fundamental limitations and real-world implementation.

2) We investigate the impact of pulse-shaping on the \ac{AFDM} signal, identifying the pulse with \ac{RRC} spectrum as the most suitable choice. We derive the analytical expression of the \ac{PSD} of \ac{AFDM}, evaluating it under realistic conditions such as pulse truncation. Furthermore, we provide a spectral comparison with other multicarrier schemes, highlighting how pulse-shaping and subcarrier suppression are necessary to maintain the multicarrier signal structure and control \ac{OOB} emissions. Despite the use of \acp{SC}, \ac{AFDM} still achieves a higher throughput than \ac{OFDM} in high-mobility channels, thanks to reduced pilot overhead and improved resilience to \ac{ICI}, confirming the robustness of \ac{AFDM} reported in the literature.

3) We extend the \ac{CT} model to analyze the impact of realistic \acp{HWI}, \ac{PN}, \ac{CFO}, and \ac{SJ}, which are often neglected in matrix-based \ac{DT} models.

4) We derive the \acp{CRB} for delay and Doppler estimation in closed form, providing a theoretical benchmark for \ac{AFDM} sensing performance in \ac{DS} channels.


The rest of the manuscript is organized as follows. Section \ref{Sec:Signal_model} introduces the fundamental \ac{CT} derivation of the \ac{AFDM} waveform based on the \ac{AFS}. In Section \ref{Sec:psd_hwi_ber_crb} the analytical \ac{PSD} is derived and the impact of \acp{HWI} on the \ac{AFDM} received signal is assessed; moreover, a theoretical \ac{BER} analysis, and the derivation of the closed-form \acp{CRB} for delay and Doppler estimation in \ac{DS} channels are provided. In Section \ref{Sec:Num_res} various numerical results and performance benchmarks to validate our analysis of \ac{AFDM} are analyzed. Finally, Section \ref{Sec:Concl} concludes the paper.

\noindent \emph{Notation}: Throughout this paper, superscripts $(\cdot)^{\ast}$ and $(\cdot)^{H}$ denote the complex conjugate and conjugate transpose (Hermitian), respectively. The operator $\mathrm{mod}_B[\cdot]$ indicates the \emph{modulo} $B$ operation, while $\ast$ denotes the \emph{linear convolution} between two signals or functions. For a complex variable $x$, we denote its \emph{real} and \emph{imaginary parts} as $\Re\{x\}$ and $\Im\{x\}$, respectively. We define a matrix $\mathbf{X}\triangleq[x_{m,n}]$, where $x_{m,n}$ represents the element in the $m$th row and $n$th column. $\mathbf{I}_N$ denotes the \emph{identity matrix} of order $N$, and $\boldsymbol{\Xi}_V$ represents the unitary \ac{DFT} matrix of order $V$, where the $(p,q)$th entry is given by $\exp(-j2\pi pq/V)/\sqrt{V}$. Finally, $\mathrm{diag}(\mathbf{x}_N)$ denotes an $N\times N$ diagonal matrix with the elements of the $N$-dimensional vector $\mathbf{x}_N$ on its main diagonal.

\section{System and Signal Models}\label{Sec:Signal_model}
In this section, we develop a \ac{CT} analytical model for \ac{AFDM}, starting from its \ac{DT} formulation. The proposed derivation follows the same approach adopted for \ac{OFDM} in \cite[Sec. II-A]{mirabella_use_2024}, but extends it to the \ac{AFD}, thus providing a rigorous framework for physical interpretation, and for investigating pulse-shaping effects and spectral properties of \ac{AFDM} signals.

The remaining part of this section is divided into three subsections. The first subsection is devoted to the generation of the \ac{AFDM} \ac{TX} signal, highlighting the role of the modulation parameters and the \ac{TX} pulse in shaping the chirp-periodic waveform.
The second subsection derives the received signal model under ideal, non-distorting channel conditions, illustrating the basic demodulation mechanism and its relationship with the \ac{DAFT}.
Finally, the third section extends the analysis to the case of a \ac{DS} channel, characterized by both time and frequency dispersion.

\subsection{Signal Generation} \label{subsec:tx_afdm_signal}

Consider the transmission of an $N$-dimensional\footnote{An even value of $N$ is assumed in this paper.} information vector $\mathbf{c}_N \triangleq [c_0,c_1,...,c_{N-1}]^T$, whose entries are symbols drawn from an $M_{\mathrm{c}}$ary constellation. This vector represents one \ac{AFDM} symbol and carries the information in the \ac{AFD}. According to \cite{bemani_affine_2023}, the mapping between the \ac{AFD} and the \ac{TD} is achieved through the \ac{IDAFT} with parameters $(\lambda_1,\lambda_2)$ as
\begin{equation}
\mathbf{x}_N = \mathbf{A}_{\lambda_1,\lambda_2}^{\mathrm{H}}\,\mathbf{c}_N \text{,}
\label{eq:IDAFT_vector}
\end{equation}
where
\begin{equation}
\mathbf{A}_{\lambda_1,\lambda_2} \triangleq \boldsymbol{\Lambda}_{\lambda_2} \, \boldsymbol{\Xi}_N \, \boldsymbol{\Lambda}_{\lambda_1}
\label{eq:DAFT_matrix}
\end{equation}
is the \ac{DAFT} matrix, and
\begin{equation}
\boldsymbol{\Lambda}_{\lambda} \triangleq \mathrm{diag}\bigl(\mathbf{a}_{N,\lambda}\bigr) \text{;}
\end{equation}
with $\mathbf{a}_{N,\lambda} \triangleq [a_{\lambda,0},...,a_{\lambda,N-1}]^T$ and $a_{\lambda,n} \triangleq \exp(-j2\pi \lambda n^2)$.

It is easy to show that the $n$th element of $\mathbf{x}_N$, in \eqref{eq:IDAFT_vector}, can be expressed as
\begin{equation}
x_n = \frac{1}{\sqrt{N}} \sum_{m=0}^{N-1} c_m \, \exp\bigl(j2\pi(\lambda_1 n^2+\lambda_2 m^2+mn/N)\bigr) \text{,}
\label{eq:IDAFT_element}
\end{equation}
and that the \emph{chirp periodicity} property
\begin{equation}
x_{n+lN} = x_n \exp\Bigl(j2\pi \lambda_1 \bigl(l^2N^2 + 2nlN\bigr)\Bigr) \text{,}
\label{eq:IDAFT_coeff_property}
\end{equation}
holds for any choice of indices $(l,n)$. Moreover, the \emph{inverse chirp periodicity} property
\begin{equation}
c_{m+lN} = c_m \exp\Bigl(-j2\pi \lambda_2 \bigl(l^2N^2 + 2mlN\bigr)\Bigr) \text{,}
\label{eq:DAFT_coeff_property}
\end{equation}
holds, for any pair of indices $(m,l)$.

In order to obtain a \ac{CT} waveform, the vector $\mathbf{x}_N$ in \eqref{eq:IDAFT_vector} is periodically extended, yielding the chirp-periodic sequence ${\bar{x}_k}$, so that $\bar{x}_k = x_k$ for any $k \in \{0,1,...,N-1\}$ and
\begin{equation}
\bar{x}_k = x_{\mathrm{mod}_N[k]} \, \exp\bigl(j2\pi\lambda_1(N^2+2Nk)\bigr),
\label{eq:chirp_periodic_element}
\end{equation}
holds for any $k\notin \{0,1,...,N-1\}$.

\begin{figure}
\centering
\includegraphics[width=0.9\columnwidth]{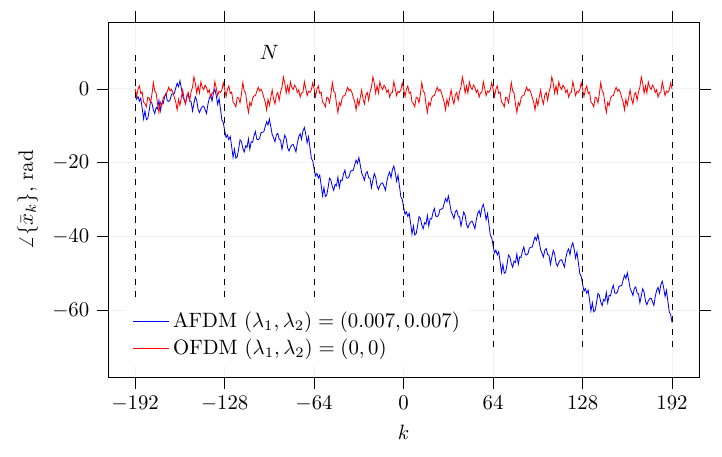}
\caption{Unwrapped phase of the \ac{DT} sequence $\{\bar{x}_k\}$ for \ac{AFDM} and \ac{OFDM}, generated on the basis of the same $4$-\ac{QAM} symbols. The vertical dashed lines indicate multiples of the symbol period $N=64$. \ac{AFDM} parameters $\lambda_1 = \lambda_2=0.007$ are used for the chirp-periodic sequence. The \ac{OFDM} curve (corresponding to the choice $\lambda_1 = \lambda_2 = 0$) is included for comparison.}
\label{fig:phase_xbar_k}
\end{figure}

As shown in Fig. \ref{fig:phase_xbar_k}, referring to the \ac{DT} chirp-periodic extension of the original vector $\mathbf{x}_N$, the phase $\angle\{\bar{x}_k\}$ of the \ac{AFDM} sequence repeats every $N$ samples. In that figure, the same sequence obtained on the basis of the same \ac{QAM} symbols with $\lambda_1 = \lambda_2 = 0$ (i.e., when \ac{OFDM} is considered) is also shown as a reference in order to highlight the effect of the chirp-periodic modulation on the phase evolution of $\mathbf{x}_N$.

The sequence $\{\bar{x}_k\}$ feeds a pulse amplitude modulator, which produces the baseband signal
\begin{equation}
s(t;\mathbf{c}_N)\!=\!\!\!\! \sum_{k=-\infty}^{+\infty} \bar{x}_k \, p(t-kT_{\mathrm{s}})
\exp({-j2\pi\lambda_1 k^2})\exp\Bigl(j2\pi\bar{\lambda}_1 t^2\Bigr) \text{,}
\label{eq:AFDM_signal_st1}
\end{equation}
where $\bar{\lambda}_1 = \lambda_1/T_{\mathrm{s}}^2$ is the chirp parameter in $\text{Hz}^2$.

The last formula deserves the following comments: 1) the resulting signal is chirp-periodic with period $T \triangleq N T_{\mathrm{s}}$, since
\begin{equation}
s(t+T;\mathbf{c}_N) = s(t;\mathbf{c}_N) \exp\bigl(j2\pi\bar{\lambda}_1(T^2+2Tt)\bigr) \text{,}
\end{equation}
and 2) the quadratic phase terms in \eqref{eq:AFDM_signal_st1} induce the time-varying frequency characteristic of \ac{AFDM} signals.

\begin{figure*}
\centering
\includegraphics[width=0.9\textwidth]{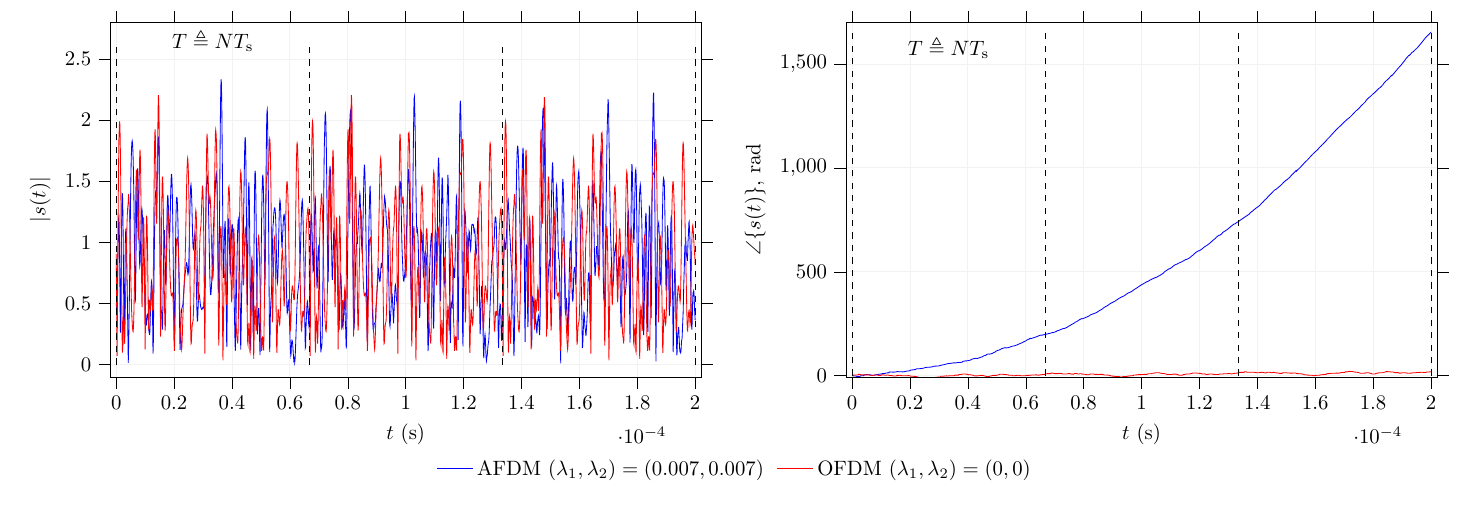}
\vspace{-3ex}
\caption{Time evolution of the magnitude (left) and unwrapped phase (right) of the \ac{CT} \ac{AFDM} (with $\lambda_1 = \lambda_2=0.007$) and \ac{OFDM} signals originating from the same sequence of $4$-\ac{QAM} channel symbols. The vertical dashed lines identify $4$ distinct instants, all multiples of the signal period $T$. The following parameters have been employed in the generation of the \ac{CT} signals: 1) $N=64$; 2) \ac{RRC} pulse $p(t)$ with roll-off factor $\alpha=0.15$; 3) oversampling factor $N_{\mathrm{c}}=10$; 4) $T_{\mathrm{s}}=1.04$ \textmu s.}
\label{fig:st_cpp}
\end{figure*}

Fig. \ref{fig:st_cpp} illustrates the magnitude and phase of the \ac{CT} \ac{AFDM} signal associated with a specific realization of the chirp-periodic sequence $\{\bar{x}_k\}$ (a $4$-\ac{QAM} is assumed). The magnitude and phase of the \ac{OFDM} signal, resulting from the choice $\lambda_1 = \lambda_2 = 0$ and originating from the same sequence of channels symbols, are also provided as a reference. Moreover, in the same figure, four vertical dashed lines are drawn at multiples of the period $T$ to highlight the chirp-periodicity of the \ac{AFDM} signal. As expected from \eqref{eq:AFDM_signal_st1}, the \ac{AFDM} phase evolves quadratically within each period, showing the time-varying frequency characteristic induced by the quadratic phase terms, whereas the \ac{OFDM} phase remains piecewise constant over each symbol period.

Because of its chirp-periodicity, which is obtained, in practice, through the insertion of a \ac{CPP} of size $N_{\mathrm{cpp}}$ \cite{bemani_affine_2023}, the signal in \eqref{eq:AFDM_signal_st1} admits the \ac{AFS} representation (see App. \ref{APP:AFS_AFT})
\begin{equation}
s(t;\mathbf{c}_N) = \sum_{m=-\infty}^{+\infty} S_{m}(\mathbf{c}_N) \, \exp\left(j 2 \pi \left(\bar{\lambda}_1 t^2 + \frac{m}{T} t\right)\right) \text{,}
\label{eq:st_cn_cpp_FS1}
\end{equation}
holding in the interval $[0, T]$; here
\begin{equation}
S_{m}(\mathbf{c}_N) = \frac{1}{T} \int_0^T\!\!\! s(t,\mathbf{c}_N) \, \exp\left(-j 2 \pi \left(\bar{\lambda}_1 t^2 + \frac{m}{T} t\right)\right) \, {\rm d}t
\label{eq:AFDM_AFS_coeff}
\end{equation}
is the $m$th coefficient of such a series; note that this coefficient depends on both the information vector $\mathbf{c}_N$ and the selected pulse shape.

Substituting the \ac{RHS} of \eqref{eq:AFDM_signal_st1} into that of \eqref{eq:AFDM_AFS_coeff}, reorganizing the summations, and exploiting the property in \eqref{eq:IDAFT_coeff_property}, one obtains, after some manipulations,
\begin{equation}
S_m(\mathbf{c}_N) = \frac{1}{\sqrt{N}T_{\mathrm{s}}}\, c_m \, P_m \, \exp\bigl(j2\pi\lambda_2 m^2\bigr) \text{,}
\label{eq:AFDM_AFS_coeff_final}
\end{equation}
where $P_m \triangleq P(m\Delta_f)$, $\Delta_f \triangleq 1/(N T_{\mathrm{s}})$ is the \emph{subcarrier spacing} and $P(f)$ denotes the \ac{CFT} of the pulse $p(t)$.

Then, substituting \eqref{eq:AFDM_AFS_coeff_final} in \eqref{eq:st_cn_cpp_FS1} yields the \ac{CT} \ac{AFDM} signal model
\vspace{-1em}
\begin{multline}
s(t;\mathbf{c}_N) = \frac{1}{\sqrt{N}T_{\mathrm{s}}} \sum_{m=-\infty}^{+\infty} c_m \, P_m \, \exp(j2\pi\lambda_2 m^2) \\
\cdot \exp\left(j 2 \pi \left(\bar{\lambda}_1 t^2 + \frac{m}{T} t\right)\right) \text{.}
\label{eq:AFDM_CT_model}
\end{multline}

If we define $m = q + kN$ with $q \in [-N/2,\, N/2-1]$, for any $k$ and using \eqref{eq:DAFT_coeff_property}, \eqref{eq:AFDM_CT_model} can be rewritten as
\begin{eqnarray}
s(t;\mathbf{c}_N) = \frac{1}{\sqrt{N}\,T_{\mathrm{s}}} \sum_{q=-N/2}^{N/2} c_{\mathrm{mod}_N[q]} \exp\Bigl(j2\pi \lambda_2 q^2\Bigr)&&
\label{eq:AFDM_pulse_expanded} \\[-1ex]
&&\hspace{-45ex} \cdot \exp\Bigl(j2\pi \Bigl(\bar{\lambda}_1 t^2 + \frac{q}{T} t\Bigr)\Bigr)\sum_{k=-\infty}^{+\infty} P_{q+kN} \, \exp\Bigl(j2\pi \frac{k}{T_{\mathrm{s}}} t\Bigr) \nonumber \text{.}\label{eq:AFDM_pulse_expanded_2}
\end{eqnarray}

As in conventional \ac{OFDM} \cite[Sec. 3.7.2]{vitetta_wireless_2013}, a \emph{strictly bandlimited} \ac{TX} pulse $p(t)$ needs to be adopted if we want that the signal $s(t;\mathbf{c}_N)$ in \eqref{eq:AFDM_pulse_expanded} takes on a multicarrier structure; in fact, this choice guarantees that only one of the addends of the sum over $k$ appearing in the \ac{RHS} of \eqref{eq:AFDM_pulse_expanded_2} is different from zero, so that \ac{SI} is prevented.
In practice, a \ac{RRC} pulse can be selected. Owing to its \emph{excess bandwidth}, only a subset $N_{\mathrm{u}} < N$ of subcarriers can be used without \ac{SI}. The remaining $N_{\mathrm{sc}} = N - N_{\mathrm{u}}$ subcarriers are therefore intentionally deactivated; for this reason, they are called \acp{SC}.

If the roll-off factor is $\alpha$, the number of active subcarriers is $N_{\mathrm{u}} = 2N_{\alpha} + 1$, with $N_{\alpha} \triangleq \lfloor N(1-\alpha)/2 \rfloor$. Therefore, the finite and infinite sums appearing in the \ac{RHS} of \eqref{eq:AFDM_pulse_expanded} involve only $N_{\mathrm{u}}$ useful terms and a single term, respectively, so that \eqref{eq:AFDM_pulse_expanded} can be expressed as
\begin{eqnarray}
s\bigl(t;\mathbf{c}_N\bigr) = \frac{1}{\sqrt{N}T_{\mathrm{s}}} \sum_{m=-N_{\alpha}}^{N_{\alpha}} c_{\mathrm{mod}_N[m]} \, P_{\mathrm{mod}_N[m]}&&
\label{eq:AFDM_rrc_pulse}\\[-0.5ex]
&&\hspace{-30ex}\cdot \exp\Bigl(j2\pi \lambda_2 m^2\Bigr) \exp\Bigl(j2\pi \Bigl(\bar{\lambda}_1 t^2 + \frac{m}{T} t\Bigr)\Bigr) \text{.}\nonumber
\end{eqnarray}

\subsection{Received Signal Model in the Presence of an Ideal Communication Channel} \label{subsec:rx_afdm_signal_ideal}

In this subsection we assume that the signal $s(t;\mathbf{c}_N)$, in \eqref{eq:AFDM_rrc_pulse}, is sent over an \emph{ideal} communication channel (i.e., a noiseless and distortionless channel), so that the signal available at the receiver is $z(t;\mathbf{c}_N) = s(t;\mathbf{c}_N)$.

This signal feeds the \ac{RX} filter, whose \ac{IR} is $g(t) = u(-t)\,\exp(j2\pi \bar{\lambda}_1 t^2)$, where $u(t)$ is a real signal.
Then, the response $r(t; \mathbf{c}_N) \triangleq z(t;\mathbf{c}_N) \ast g(t)$, becomes
\begin{eqnarray}
r(t; \mathbf{c}_N) = \frac{1}{\sqrt{N}T_{\mathrm{s}}} \sum_{m=-N_{\alpha}}^{N_{\alpha}} c_{\mathrm{mod}_N[m]} \, P_{\mathrm{mod}_N[m]}&&
\label{eq:rt_ideal}\\ 
&&\hspace{-42ex}\cdot U^{\ast}\Bigl(2\bar{\lambda}_1 t + \frac{m}{NT_{\mathrm{s}}}\Bigr) \,\exp\left(j 2 \pi \left(\bar{\lambda}_1 t^2 + \frac{m}{T} t + \lambda_2 m^2\right)\right) \text{,}\nonumber
\end{eqnarray}
with $U(f)$ being the \ac{CFT} of $u(t)$.

As it can be inferred from the last result, \ac{RX} filtering results into the introduction of a factor $U(\cdot)^{\ast}$, sampled at a time-dependent instant and also depending on the index $m$. Again, note that, when $\lambda_1 = 0$ (i.e., $\bar{\lambda}_1=0$), $U(\cdot)$ is sampled at the fixed frequency $f_m = m/T$ (the same occurs for \ac{OFDM}).

The signal $r(t; \mathbf{c}_N)$ \eqref{eq:rt_ideal} is sampled at the instants $t_{\tilde{n}} \triangleq \tilde{n}T_{\mathrm{s}}$ (with $\tilde{n}=0,1,...,N + N_{\mathrm{cpp}}-1$). It is easy to show that the $\tilde{n}$th sample $\breve{r}_{\tilde{n}} \triangleq r(t_{\tilde{n}}; \mathbf{c}_N)$, after removing the \ac{CPP} (i.e., $\tilde{n}=0,1,...,N-1$), can be expressed as
\vspace{-2ex}
\begin{multline}
\!\!\!\breve{r}_{\tilde{n}} = \frac{1}{\sqrt{N}T_{\mathrm{s}}}\!\sum_{m=-N_{\alpha}}^{N_{\alpha}}\!\!\! c_{\mathrm{mod}_N[m]} \, P_{\mathrm{mod}_N[m]}U^{\ast}\left(2\frac{\lambda_1}{T_{\mathrm{s}}}\tilde{n} + \frac{m}{NT_{\mathrm{s}}}\right)\\
\cdot \exp\left(j 2 \pi \left(\lambda_1 \tilde{n}^2 + \frac{m}{N} \tilde{n} + \lambda_2 m^2\right)\right) \text{.}
\label{eq:r_tilden_ideal1}
\end{multline}

This last result shows that, unlike \ac{OFDM}, in \ac{AFDM} the impact of the \ac{RX} pulse depends also on the sampling instant through $\tilde{n}$.
In order to proceed further in the derivations we need to select the normalized chirp parameter and the \ac{RX} pulse bandwidth properly.
Regarding the first parameter, following \cite{bemani_affine_2023}, we select $\lambda_1 = A/(2N)$, with $A \in \mathbb{Z}$.
As far as the \ac{RX} pulse is concerned, its spectrum is required to be \emph{strictly bandlimited}. In particular we set
\begin{equation}
U^{\ast}\bigl(f_{\tilde{n},m}\bigr) = \sqrt{T_s} \, \mathrm{rect}\left(\frac{f_{\tilde{n},m}}{B_{\mathrm{rx}}}\right),
\label{eq:rx_pulse_spectrum}
\end{equation}
where $\mathrm{rect}(x)$ is a rectangular function defined as $1$ for $|x|\leq 1/2$ and $0$ otherwise, $f_{\tilde{n},m} \triangleq (A\tilde{n} + m)\Delta_f$ and $B_{\mathrm{rx}}$ is the \ac{RX} pulse bandwidth.

By choosing $A=1$ and $B_{\mathrm{rx}}=\lceil 2N(1+\alpha)\rceil\Delta_f$, it follows that
$U^{\ast}(f_{\tilde{n},m})=\sqrt{T_{\mathrm{s}}}$ for all 
$\tilde{n}\in\{0,\ldots,N-1\}$ and 
$m\in\{-N_{\alpha},\ldots,N_{\alpha}\}$.
These selections allow us to rewrite \eqref{eq:r_tilden_ideal1} as
\begin{eqnarray}
\breve{r}_{\tilde{n}} \triangleq r(t_{\tilde{n}}; \mathbf{c}_N) = \frac{1}{\sqrt{N}} \sum_{m=-N_{\alpha}}^{N_{\alpha}} c_{\mathrm{mod}_N[m]}&&
\label{eq:r_tilden_ideal2}\\[-1ex]
&&\hspace{-20ex}\cdot\exp\bigl(j 2 \pi \bigl(\lambda_1 \tilde{n}^2 + \frac{m}{N} \tilde{n} + \lambda_2 m^2\bigr)\bigr) \nonumber \text{,}
\end{eqnarray}
provided that $P_{\mathrm{mod}_N[m]} = \sqrt{T_{\mathrm{s}}}$ for each considered $m$.

Storing the sequence $\{\breve{r}_{\tilde{n}};\tilde{n}=0,1,...,N-1\}$ into the $N$-dimensional vector
\begin{equation}
\mathbf{r}_N \triangleq [\breve{r}_0,\breve{r}_1,...,\breve{r}_{N-1}]^T = \mathbf{A}_{\lambda_1,\lambda_2}^H \, \mathbf{c}_N \text{,}
\label{eq:rN_vector_ideal}
\end{equation}
leads us to the conclusion that $\mathbf{c}_N$ can be recovered from $\mathbf{r}_N$ by applying an order $N$ \ac{DAFT}, with parameters $(\lambda_1,\lambda_2)$, to the vector $\mathbf{r}_N$. In fact, we have that
\begin{equation}
\mathbf{y}_N \triangleq [y_0,y_1,...,y_{N-1}]^T = \mathbf{A}_{\lambda_1,\lambda_2} \, \mathbf{r}_N = \mathbf{c}_N \text{.}
\end{equation}

\subsection{Received Signal Model in the Presence of a DS Channel}
\label{subsec:rx_afdm_signal_ds}
Let us focus now on the reception on a \ac{LTV} multipath fading channel characterized by the \ac{CIR}
\begin{equation}
h(t,\tau) = \sum_{l=0}^{L-1} h_l(t,\tau) \text{,}
\label{eq:ht_tau}
\end{equation}
where
\begin{equation}
h_l(t,\tau) \triangleq a_l\,\exp(j2\pi \nu_l t)\,\delta(\tau-\tau_l)
\label{eq:hlt_tau}
\end{equation}
denotes the $l$th path contribution, while $a_l$, $\tau_l$ and $\nu_l$ represent the gain, the delay \footnote{We assume the \ac{CIR} components are arranged in ascending order of delays, so that $\tau_0$ and $\tau_{L-1}$ are the minimum and maximum delays, respectively.} and the Doppler shift of the $l$th channel path, respectively, and $L$ is the overall number of paths.

We first consider the contribution of the $l$th channel path $h_l(t,\tau)$ (see \eqref{eq:hlt_tau}) to the \ac{RX} signal after \ac{RX} filtering, that is
\begin{equation}
r_l(t; \mathbf{c}_N) \triangleq s(t;\mathbf{c}_N) \ast h_{l}(t,\tau) \ast g(t) \text{,}
\label{r_l_component}
\end{equation}
where the additive noise term has been neglected for simplicity. It is not difficult to prove that the last signal can be alternatively expressed as
\begin{eqnarray}
\hspace{-3ex}r_l(t; \mathbf{c}_N) = \frac{\tilde{a}_l}{\sqrt{N}T_{\mathrm{s}}}\!\sum_{m=-N_{\alpha}}^{N_{\alpha}} \!\!\!\!c_{\mathrm{mod}_N[m]} \, P_{\mathrm{mod}_N[m]}\exp\bigl(j2\pi \lambda_2 m^2\bigr)\hspace{-8ex}&& \nonumber \\
&&\hspace{-48ex}\cdot \exp\Bigl(-j2\pi \frac{m}{T} \tau_l \Bigr) U^{\ast}\Bigl(2\bar{\lambda}_1 (t-\tau_l) + \frac{m}{T}+\nu_l\Bigr)\nonumber \\
&&\hspace{-48ex}\cdot \exp\left(j 2 \pi \left(\bar{\lambda}_1 t^2 + \frac{m}{T} t\right)\right) \, \exp\bigl(j2\pi (\nu_l - 2\bar{\lambda}_1 \tau_l) \,t\bigr) \text{,}
\label{eq:rlt_ds_chann}
\end{eqnarray}
where $\tilde{a}_l \triangleq a_l \exp(j2\pi \bar{\lambda}_1 \tau_l^2) \exp(-j2\pi \nu_l \tau_l)$.

After 1) sampling the last signal at the instants $t_{\tilde{n}} \triangleq \tau_{L-1} + \tilde{n}T_{\mathrm{s}}$ and 2) following the same procedure employed for the derivation of \eqref{eq:r_tilden_ideal2}, one obtains
\begin{multline}
\bar{r}_{l,\tilde{n}} = \frac{\breve{a}_l}{\sqrt{N}T_{\mathrm{s}}} \sum_{m=-N_{\alpha}}^{N_{\alpha}} c_{\mathrm{mod}_N[m]} \, P_{\mathrm{mod}_N[m]} \, \bar{U}_{m}^{\ast} \\  
\cdot \exp\left(j2\pi \left(\lambda_1 \tilde{n}^2 + \frac{m}{N}\tilde{n} + \lambda_2 m^2\right)\right) \\
\cdot \exp\bigl(j2\pi m F_{\tau_l}\bigr) \, \exp\Bigl(j2\pi (F_{\nu_l}/N + 2\lambda_1 N F_{\tau_l})\tilde{n}\Bigr) \text{,}
\label{eq:rln1_simpl}
\end{multline}
where
\begin{equation}
\breve{a}_l \triangleq \tilde{a}_l \exp(j2\pi\bar{\lambda}_1\tau_{L-1}^2) \, \exp\bigl(j2\pi(\nu_l-2\bar{\lambda}_1\tau_l)\,\tau_{L-1}\bigr) \text{,}
\end{equation}
\begin{equation}
F_{\tau_l} \triangleq (\tau_{L-1}-\tau_l)/(NT_{\mathrm{s}}) \text{,}
\end{equation}
and
\begin{equation}
F_{\nu_l} \triangleq \nu_l / \Delta_f
\end{equation}
respectively denote the complex gain, the normalized delay and the normalized Doppler frequency associated\footnote{Note that the derivation of \eqref{eq:rln1_simpl} relies on the assumption that $|NF_{\tau_l}|< N$ and $|NF_{\nu_l}|< N$ $\forall l$.} with the $l$th channel path.

Given \eqref{eq:rln1_simpl}, we can account for all the channel paths by summing $\bar{r}_{l,\tilde{n}}$ \eqref{eq:rln1_simpl} over $l$, i.e., by defining
\begin{equation}
\breve{r}_{\tilde{n}} \triangleq \sum_{l=0}^{L-1} \bar{r}_{l,\tilde{n}} \text{.}
\end{equation}

Then, the sequence $\{\breve{r}_{\tilde{n}};\tilde{n}=0,1,...,N-1\}$ is stored into the $N$-dimensional vector $\mathbf{r}_N \triangleq [\breve{r}_0,\breve{r}_1,...,\breve{r}_{N-1}]^T$ and undergoes order $N$ \ac{DAFT} processing, with parameters $(\lambda_1,\lambda_2)$; this yields
\begin{equation}
\mathbf{y}_N \triangleq [y_0, y_1,...,y_{N-1}]^T = \mathbf{A}_{\lambda_1,\lambda_2}\,\mathbf{r}_N \text{.}
\label{eq:yN_vector1}
\end{equation}

The last vector can be expressed, including an \ac{AWGN} contribution, as
\begin{equation}
\mathbf{y}_N = \mathbf{H} \,\mathbf{c}_N + \mathbf{w}_N\text{,}
\label{eq:yN_vector2}
\end{equation}
where $\mathbf{H} \triangleq [H_{m,n}]$ denotes the $N\times N$ \emph{effective channel matrix} and $\mathbf{w}_N \triangleq [w_0, w_1,...,w_{N-1}]^T$ is the \ac{AWGN} vector affecting $\mathbf{y}_N$. Moreover, the $(m,n)$th coefficient of the effective channel matrix $\mathbf{H}$, in \eqref{eq:yN_vector2}, can be expressed as
\begin{eqnarray}
\label{eq:Hmn_doubly_selective1}
H_{m,n} = \frac{1}{T_{\mathrm{s}}} P_m \, \bar{U}_{m}^{\ast} \exp\bigl(j2\pi \lambda_2 (m^2 - n^2)\bigr)&& \\[-1ex]
&&\hspace{-20ex}\cdot \sum_{l=0}^{L-1}\breve{a}_l \, \exp\bigl(j2\pi m F_{\tau_l}\bigr) \,\mathcal{F}_l[n-m] \text{,}
\nonumber
\end{eqnarray}
where
\begin{equation}
\mathcal{F}_l[x] = G_N\bigl(F_{\nu_l}/N + 2\lambda_1 N F_{\tau_l} - {x}/{N}\bigr)
\end{equation}
is the term describing the nature of \ac{ICI} in \ac{AFDM} modulation (for a given index $x$), whereas
\begin{equation}
G_X(\theta) \triangleq \frac{1}{X} \frac{\exp(-j2\pi\theta X)-1}{\exp(-j2\pi\theta)-1}
\label{eq:dirichlet_kernel}
\end{equation}
is the \emph{Dirichlet kernel} of order $X$ and phase $\theta$. A characterization of the channel matrix $\mathbf{H}$ is provided in App. \ref{App:AFD_IR}.

\section{Power spectral density, receiver impairments, and fundamental bounds of AFDM}
\label{Sec:psd_hwi_ber_crb}

In this section, the \ac{CT} developed in the previous section is exploited to investigate various technical issues related to the adoption of \ac{AFDM} in communication \& sensing applications. In particular, we first evaluate the \ac{PSD} of the \ac{TX} \ac{AFDM} signal to analyze its spectral properties and unveil their dependence on the chirp parameters and pulse-shaping.
Then, we investigate the impact of the three different receiver impairments listed in Tab. \ref{tab:hardware_impairments} (namely, \ac{PN}, \ac{CFO}, and \ac{SJ}) on communication performance.
An evaluation of the theoretical \ac{BER} achieved with a \ac{LMMSE} receiver is then provided; this offers further insights into the robustness of \ac{AFDM} in the presence of \acp{HWI} and \ac{DS} channels.
Finally, we derive the \ac{CRB} for the estimation of channel delay and Doppler shifts; our results provide fundamental limits useful to assess \ac{AFDM} sensing performance.
The part complements our results about the use of \ac{AFDM} in digital communication. In fact, it allows us to compare channel estimation accuracy in an \ac{AFDM}-based transmission with that achievable with \ac{OFDM}. Moreover, it shows how the chirp parameter $\lambda_1$ influences the derived theoretical bounds.

\vspace{-2ex}
\subsection{Power Spectral Density of TX AFDM Signals}

In this subsection, the derivation of the \ac{PSD} for \ac{AFDM} modulation is sketched. To begin, we rewrite the signal $s(t;\mathbf{c}_N)$ \eqref{eq:AFDM_signal_st1} as
\begin{equation}
s\bigl(t;\mathbf{c}_N\bigr)\! = \!\!\!\!\!\sum_{k=-N_{\mathrm{cpp}}}^{N-1} \!\!\!\!\!\bar{x}_k \, p(t-kT_{\mathrm{s}}) \exp\bigl(j2\pi \bigl(\bar{\lambda}_1 t^2 - \lambda_1 k^2 \bigr)\bigr) \text{,}
\label{eq:st_cn_cpp_finite}
\end{equation}
where the infinite sum appearing in the \ac{RHS} of \eqref{eq:AFDM_signal_st1} has been replaced by a finite counterpart. Based on the last model of the complex envelope of the transmitted signal, the approach illustrated in \cite[Sec. 3.7.3]{vitetta_wireless_2013} can be adopted to derive the \ac{PSD} of \ac{AFDM}. For this reason, we first evaluate the \emph{average} \ac{PSD} of the data sequence $\{x_k\}$. Then, we apply \cite[Eq. (3.66)]{vitetta_wireless_2013}, expressing the \ac{PSD} of a \ac{PAM} signal, conveying a cyclostationary data sequence.

In our derivations we assume that the $N_{\mathrm{u}} = 2N_{\alpha}+1$ useful elements of the channel symbol vector $\mathbf{c}_N$ have the following properties: 1) they belong to an $M_{\mathrm{c}}$ary constellation; 2) they are statistically independent and identically distributed (i.i.d.); 3) they have zero mean and variance $\sigma^2_{\mathrm{c}}$.
The last two assumptions imply that the elements of sequence $\{\bar{x}_k\}$ are also i.i.d., thanks to the orthogonality property of the \ac{IDAFT} transform in \eqref{eq:IDAFT_vector}; moreover, they have zero mean and the same variance.
Based on \cite[Eq. (3.287)]{vitetta_wireless_2013}, it is easy to show that the average \ac{PSD} of the sequence $\{\bar{x}_k\}$ is
\begin{equation}
\bar{S}_{\mathrm{x}}(f) = \frac{\sigma^2_{\mathrm{c}}}{N \, N_{\mathrm{T}}} \sum_{l=-N_{\alpha}}^{N_{\alpha}} \frac{\sin^2\bigl(\pi N_{\mathrm{T}}(f-f_l) T_{\mathrm{s}}\bigr)}{\sin^2\bigl(\pi(f-f_l) T_{\mathrm{s}}\bigr)} \text{,}
\label{PSD_x_overbar}
\end{equation}
where $N_{\mathrm{T}}\triangleq N+N_{\mathrm{cpp}}$ and $f_l = l/(NT_{\mathrm{s}})$ is the frequency of the $l$th subcarrier of the baseband \ac{AFDM} signal.
\vspace{-3ex}
\begin{table}[H]
\centering
\caption{Receiver impairments considered in our \ac{AFDM} analysis.}
\label{tab:hardware_impairments}
\vspace{-1ex}
\begin{tabular}{lll}
\hline
\textbf{Impairment}  & \textbf{References} & \textbf{System model} \\
\hline
Phase Noise &  \cite{bemani_affine_2021,sui_mimo_2026} & \ac{DT} \\
Carrier Frequency Offset  & \cite{bemani_affine_2021,sui_mimo_2026} & \ac{DT} \\
Sampling Jitter & - & - \\
\hline
\end{tabular}
\end{table}
\vspace{-1em}
\begin{figure}[H]
\centering
\includegraphics[width=0.9\columnwidth]{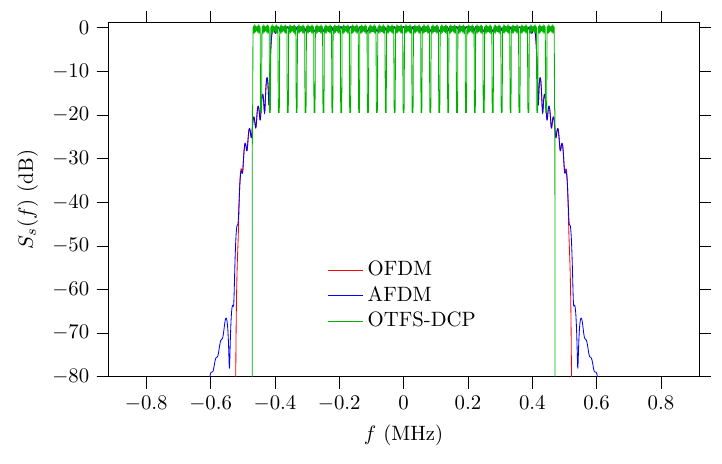}
\vspace{-3ex}
\caption{Normalized \acp{PSD} of \ac{OFDM}, \ac{AFDM} ($\lambda_1=\lambda_2=0.007$) and \ac{OTFS}-\ac{DCP}. All the considered waveforms occupy the same bandwidth $B=1$ MHz and use $N=64$ subcarriers. The size of the \ac{OTFS}-\ac{DCP} symbol matrix is $M \times N$, with $M=32$; moreover, this format employs the same ideal (untruncated) \ac{RRC} pulse $p(t)$ (with roll-off factor $\alpha=0.15$) as \ac{OFDM} and \ac{AFDM}, and FD \ac{CP} and a postfix of sizes $N_{\mathrm{cp}}^{(\mathrm{FD})}=N_{\mathrm{cpo}}^{(\mathrm{FD})}=1$, respectively.}
\label{fig:psd_modulations}
\vspace{2ex}
\includegraphics[width=0.9\columnwidth]{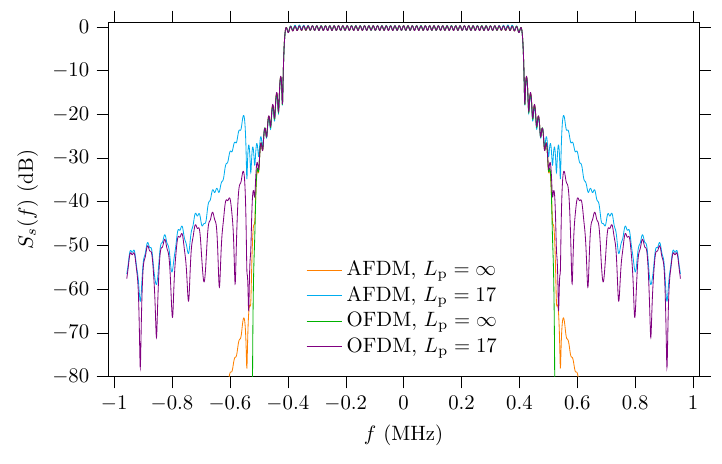}
\vspace{-3ex}
\caption{Normalized \acp{PSD} of \ac{OFDM} and \ac{AFDM} signals for ideal (untruncated) and truncated ($L_{\mathrm{p}}=17$) pulse-shaping. For \ac{AFDM}, $\lambda_1=\lambda_2=0.007$ have been assumed.}
\label{fig:ppsd_trunc_id}
\vspace{-1ex}
\end{figure}

Given \eqref{PSD_x_overbar}, the average \ac{PSD} of $s(t,\mathbf{c}_N)$ \eqref{eq:st_cn_cpp_finite} can be evaluated as
\begin{equation}
S_{\mathrm{s}}(f) = \frac{1}{T_{\mathrm{s}}} \bar{S}_{\mathrm{x}}(f) \, \bigl\vert P(f) \ast W(f) \bigr\vert^2 \text{,}
\end{equation}
where the convolution between $P(f)$ and $W(f)$ arises from the chirp modulation term introduced in \ac{AFDM}. Moreover\footnote{Note that, when $\lambda_1=0$, then $W(f) = \delta(f)$ $\forall f$, i.e., it coincides with the \emph{Dirac} delta function.}
\begin{equation}
W(f) \triangleq \frac{1}{\sqrt{\lambda_1}} \exp\Bigl(-j\pi\frac{f^2}{2\bar{\lambda}_1}\Bigr) \frac{1+j}{2} \text{.}
\end{equation}
represents the \ac{CFT} of the quadratic-phase complex chirp introduced by the \ac{AFDM}, which is solved as a Fresnel integral.

The \acp{PSD} of \ac{OFDM}, \ac{AFDM}, and \ac{OTFS}-\ac{DCP} (normalized with respect to the factor $\sigma^2_{\mathrm{c}}/(N\,N_{\mathrm{T}})$) are compared in Fig. \ref{fig:psd_modulations} compared.  These results have been obtained under the following assumptions: a) same bandwidth ($B=1$ MHz) and number of subcarriers ($N=64$) for all the considered modulation formats, b) size of the \ac{OTFS}-\ac{DCP} symbol matrix equal to $M \times N$, with $M=32$; c) same ideal \ac{RRC} pulse $p(t)$, with $\alpha=0.15$, adopted for \ac{OFDM} and \ac{AFDM}. These results lead easily to the following conclusions:

1) the \ac{AFDM} \ac{PSD} shape is different from that of \ac{OFDM}; this is due to the introduction of the chirp factor $\lambda_1$. In general, increasing $\lambda_1$ results in higher sidelobes (and, consequently, \ac{OOB} emissions).

2) The adoption of \ac{FD} \ac{CP} and postfix in  \ac{OTFS}-\ac{DCP} generates visible spectral components outside the nominal band. Note, however, these \ac{FD} additional components are canceled by the filter-bank employed at the \ac{RX} side \cite{mirabella_use_2024}.

The impact of pulse truncation on spectral containment for both \ac{OFDM} and \ac{AFDM} is illustrated in Fig. \ref{fig:ppsd_trunc_id}. When an ideal (i.e., untruncated) pulse $p(t)$ is considered, both waveforms exhibit comparable \ac{OOB} radiation. The reported \ac{OOB} levels refer to the \emph{out-of-band energy}, computed as the integral of the \ac{PSD} outside the nominal transmission bandwidth and expressed in dB with respect to the total transmitted energy. Under ideal conditions, the \ac{OOB} energy is equal to $-40$ dB and $-39.24$ dB for \ac{OFDM} and \ac{AFDM}, respectively (see Fig. \ref{fig:psd_modulations}), confirming a similar spectral confinement. When the pulse duration is truncated to $L_{\mathrm{p}} T_s$ s, with $L_{\mathrm{p}}=17$, the spectral leakage increases for both modulations. In this case, the \ac{OOB} energy rises to $-37$ dB for \ac{OFDM} and to $-30$ dB for \ac{AFDM}. Hence, \ac{OFDM} experiences a moderate degradation (approximately $3$ dB), whereas \ac{AFDM} exhibits a significantly larger increase in \ac{OOB} emissions. This indicates that \ac{AFDM} is more sensitive to pulse truncation; which can be related to the presence of an additional chirp modulation.

\subsection{Impact of Phase Noise and Carrier Frequency Offset on the Received Channel Model}
\label{subsec:AFDM_DS_PN_CFO}

In this section, we extend the received signal model developed in Subsec.  \ref{subsec:rx_afdm_signal_ds} in order to account for \ac{PN} and \ac{CFO} at the \ac{RX} side. In our analysis, we assume both \ac{PN} and \ac{CFO} vary slowly over one sampling interval (i.e., $T_{\mathrm{s}}$ s). Under the above mentioned impairments, the contribution of the $l$th component of the \ac{RX} signal filtered by the \ac{DS} channel (see \ref{r_l_component}) to the output of the \ac{RX} filter can be expressed as
\vspace{-0.5em}
\begin{multline}
r_l(t; \mathbf{c}_N)^{(\mathrm{PN+CFO})} = \Bigl(\bigl(s(t;\mathbf{c}_N) \ast h_l(t,\tau)\bigr) \\
\cdot\exp\bigl(j\phi(t)+j2\pi f_{\mathrm{cfo}}\,t\bigr)\Bigr) \ast g(t) \text{,}
\label{eq:rlt_pn_cfo}
\end{multline}
where $\phi(t)$ denotes a real-valued stochastic process\footnote{The random process $\phi(t)$ is often modeled as a slowly varying \emph{Wiener process}, whose increments are independent Gaussian random variables \cite{sui_performance_2025}.} accounting for the \ac{PN} introduced by the \ac{RX} \ac{LO}, while $f_{\mathrm{cfo}}$ represents the \ac{CFO} introduced by the receiver. Following the same line of reasoning as in Subsec.  \ref{subsec:rx_afdm_signal_ds}, it can be proved that sampling $r_l(t; \mathbf{c}_N)^{(\mathrm{PN+CFO})}$at $t_{\tilde{n}} = \tau_{L-1} + \tilde{n}T_{\mathrm{s}}$ (with $\tilde{n}=0,1,...,N-1$) yields
\begin{multline}
\bar{r}_{l,\tilde{n}} \triangleq r_l(t_{\tilde{n}}; \mathbf{c}_N) \cong \frac{\check{a}_l}{\sqrt{N}T_{\mathrm{s}}} \sum_{m=-N_{\alpha}}^{N_{\alpha}} c_{\mathrm{mod}_N[m]} \, P_{\mathrm{mod}_N[m]} \, {U}_{m}^{\ast} \\
\cdot \exp\bigl(j2\pi m F_{\tau_l}\bigr) \, \exp\Bigl(j2\pi \Bigl(\lambda_1 \tilde{n}^2 + \frac{m}{N}\tilde{n} + \lambda_2 m^2\Bigr)\Bigr) \\  
\cdot \exp\Bigl(j2\pi \bigl((F_{\nu_l}+F_{\mathrm{cfo}})/N + 2\lambda_1 NF_{\tau_l}\bigr)\tilde{n}\Bigr) \exp(j\phi_{\tilde{n}}) \text{,}
\label{eq:rln_pn_cfo}
\end{multline}
where $\check{a} = \breve{a} \exp\bigl(j2\pi f_{\mathrm{cfo}} \tau_{L-1}\bigl)$, $F_{\mathrm{cfo}} \triangleq f_{\mathrm{cfo}}/\Delta_f$ is the \ac{CFO} normalized with respect to $\Delta_f$ and $\phi_{\tilde{n}} \triangleq \phi(\tilde{n}T_{\mathrm{s}})$ represents the $\tilde{n}$th sample of the \ac{PN} process.

In order to account for the presence lo all the channel paths we sum $\bar{r}_{l,\tilde{n}}$ \eqref{eq:rln_pn_cfo} over $l$, so generating $\breve{r}_{\tilde{n}} \triangleq \sum_{l=0}^{L-1} \bar{r}_{l,\tilde{n}}$, with $\tilde{n}=0,1,...,N-1$. The resulting sequence $\{\breve{r}_{\tilde{n}};\tilde{n}=0,1,...,N-1\}$ is stored in the $N$-dimensional vector $\mathbf{r}_N \triangleq [\breve{r}_0,\breve{r}_1,...,\breve{r}_{N-1}]^T$ , which undergoes order $N$ \ac{DAFT} processing, with parameters $(\lambda_1,\lambda_2)$; this yields
\begin{equation}
\mathbf{y}_N \triangleq [y_0, y_1,...,y_{N-1}]^T = \mathbf{A}_{\lambda_1,\lambda_2}\,\mathbf{r}_N \text{.}
\label{y_n_Nonoise}
\end{equation}
Then, we include an \ac{AWGN} contribution in the \ac{RHS} of \ref{y_n_Nonoise} and adopt the linear approximation $\phi_{\tilde{n}} \cong \bar{\phi}_0 + \tilde{n}\,T_{\mathrm{s}}\bar{\phi}_1$ for the discrete time process $\phi_{\tilde{n}}$; here, $(\bar{\phi}_0,\bar{\phi}_1)$ represent the coefficients of the first‑order polynomial approximation of $\{\phi_{\tilde{n}}\}$ over the considered model. Then, \ref{y_n_Nonoise}  turns into
\begin{equation}
\mathbf{y}_N = \mathbf{H}^{(\mathrm{PN} + \mathrm{CFO})} \, \mathbf{c}_N + \mathbf{w} \text{,}
\label{eq:yN_ds_pn_cfo}
\end{equation}
where $\mathbf{w} \triangleq [w_0, w_1, ...,w_{N-1}]$ is an $N$dimensional \ac{AWGN} vector whose elements have zero mean and variance $\sigma^2_{\mathrm{w}}$, and  $\mathbf{H}^{(\mathrm{PN} + \mathrm{CFO})} \triangleq [H_{m,n}^{(\mathrm{PN} + \mathrm{CFO})}]$ represents the \emph{effective channel matrix} accounting not only for the \ac{DS} channel, but also \ac{PN} and \ac{CFO} impairments. In fact, its element $(m,n)$  can be expressed as
\begin{multline}
H_{m,n}^{(\mathrm{PN} + \mathrm{CFO})} = \frac{1}{T_{\mathrm{s}}} P_m \, U_m^{\ast} \, \exp\bigl(j2\pi \lambda_2 (m^2 - n^2)\bigr) \\
\sum_{l=0}^{L-1}\acute{a}_l \, \exp\bigl(j2\pi m F_{\tau_l}\bigr) \, \mathcal{Q}_l[n-m] \text{,}
\label{eq:Hmn_doubly_selective_pn_cfo}
\end{multline}
where $\acute{a}_l \triangleq \check{a}_l \exp(j\bar{\phi}_0)$ and (see \eqref{eq:dirichlet_kernel}) $\mathcal{Q}_l[x] = G_N((F_{\nu_l} + F_{\mathrm{cfo}})/N + 2\lambda_1 N F_{\tau_l} - {x}/{N} + {\bar{\phi}_1}/{2\pi})$ for any integer $x$.

\vspace{-2ex}
\subsection{Impact of Sampling Jitter on the Received Channel Model}
\label{subsec:AFDM_DS_SJ}

In this section, we extend the received signal model developed in Subsec.  \ref{subsec:rx_afdm_signal_ds} in order to account for \ac{SJ} at the \ac{RX} side. For this reason, we first consider again the signal $r_l(t; \mathbf{c}_N)$, in \eqref{eq:rlt_ds_chann}, assume it to be sampled at the instant
\begin{equation}
t_{\tilde{n}}^{(\mathrm{sj})} \triangleq \tau_{L-1} + \tilde{n}T_{\mathrm{s}} + \delta_{\mathrm{sj}}(\tilde{n}T_{\mathrm{s}}) \text{,}
\label{eq:sampling_instants_jitter_ds}
\end{equation}
with $\tilde{n}=0,1,\dots,N-1$; here, $\delta_{\mathrm{sj}}(\cdot)$ is a real-valued stochastic process modeling \ac{SJ} (i.e., representing a deviation from the ideal sampling grid). To simplify the following derivations we assume the linear model
\begin{equation}
\delta_{\mathrm{sj}}(\tilde{n}T_{\mathrm{s}})
= \bar{\delta}_0 + \tilde{n}T_{\mathrm{s}}\bar{\delta}_1 \text{,}
\label{eq:sj_linear_model}
\end{equation}
for $\{\tilde{n}T_{\mathrm{s}}\}$; here, $\bar{\delta}_0$ represents a constant timing offset, while $\bar{\delta}_1$ models a sampling clock skew.  Then, it is easy to show that the $\tilde{n}$th sample of $r_l(t; \mathbf{c}_N)$ can be expressed as
\vspace{-1ex}
\begin{multline}
\bar{r}_{l,\tilde{n}}^{\mathrm{(sj)}} \cong \frac{\check{a}_l}{\sqrt{N}T_{\mathrm{s}}} \sum_{m=-N_{\alpha}}^{N_{\alpha}} c_{\mathrm{mod}_N[m]} \, P_{\mathrm{mod}_N[m]}\, U^{\ast}_m \\
\cdot \exp\bigl(j2\pi m F_{\tau_l}\bigr) \exp\Bigl(j2\pi \Bigl(\lambda_1 \tilde{n}^2 +\frac{m}{N}\tilde{n} + \lambda_2 m^2\Bigr)\Bigr) \\
\cdot \exp\Bigl(j2\pi \Bigl({F_{\nu_l}^{(\mathrm{sj})}}/{N} + 2\lambda_1 N F_{\tau_l}^{(\mathrm{sj})}\Bigr)\tilde{n}\Bigr) \text{,}
\end{multline}
where $\check{a}_l \triangleq \breve{a}_l \exp(j2\pi (\nu_l - 2 \bar{\lambda}_1 \tau_l)\bar{\delta}_0)$, $F_{\nu_l}^{(\mathrm{sj})} \triangleq F_{\nu_l}(1+\bar{\delta}_1)$ and $F_{\tau_l}^{(\mathrm{sj})} \triangleq (\tau_{L-1}- \tau_l(1+\bar{\delta}_1))/(NT_{\mathrm{s}})$.

Based on the last result, we can easily develop the $N$-dimensional vector model $\mathbf{r}_N^{(\mathrm{sj})} \triangleq [\breve{r}_0^{(\mathrm{sj})}, \breve{r}_1^{(\mathrm{sj})},..., \breve{r}_{L-1}^{(\mathrm{sj})}]^T$ of the noiseless \ac{RX} signal, where $\breve{r}_{\tilde{n}}^{(\mathrm{sj})} = \sum_{l=0}^{L-1}\bar{r}_{l,\tilde{n}}^{\mathrm{(sj)}}$ for any $\tilde{n}$ (in the last formula, the sum accounts for the presence of $L$ channel paths). The vector $\mathbf{r}_N^{(\mathrm{sj})}$ undergoes an order $N$ \ac{DAFT}, with parameters $(\lambda_1,\lambda_2)$; if an \ac{AWGN} contribution is included now in the model (see \eqref{eq:yN_ds_pn_cfo}), this produces
\begin{equation}
\mathbf{y}_N^{(\mathrm{sj})} \triangleq \mathbf{A}_{\lambda_1,\lambda_2}\,\mathbf{r}_N^{(\mathrm{sj})} = \mathbf{H}^{(\mathrm{sj})} \mathbf{c}_N + \mathbf{w}_N \text{,}
\label{eq:yN_ds_sj}
\end{equation}
where $\mathbf{H}^{(\mathrm{sj})} \triangleq [H_{m,n}^{(\mathrm{sj})}]$ is the $N \times N$ effective channel matrix; the element $(m,n)$ of the last matrix can be expressed as

\begin{multline}
H_{m,n}^{(\mathrm{sj})} = \frac{1}{T_{\mathrm{s}}} P_m \, U_m^{\ast} \, \exp\bigl(j2\pi \lambda_2 (m^2 - n^2)\bigr) \\
\cdot \sum_{l=0}^{L-1}\check{a}_l \, \exp\bigl(j2\pi m F_{\tau_l}\bigr) \,\mathcal{K}_l[n-m] \text{,}
\label{eq:Hmn_doubly_selective_sj}
\end{multline}
where, for any integer $x$, $\mathcal{K}_l[x] = G_N(F_{\nu_l}^{(\mathrm{sj})}/N + 2\lambda_1 N F_{\tau_l}^{(\mathrm{sj})} - {x}/{N})$  represents the \ac{ICI} term corrupted by \ac{SJ} (see \eqref{eq:dirichlet_kernel}).

\subsection{Bit-error-rate Analysis under LMMSE Detection}

The received signal models  \eqref{eq:yN_vector2}, 
\eqref{eq:yN_ds_pn_cfo} and \eqref{eq:yN_ds_sj} developed in the previous subsections can be exploited to assess \ac{AFDM} system performance in terms of \ac{BER} when a \ac{LMMSE} detector is used at the \ac{RX} side. Given one of the above mentioned models, the \ac{LMMSE} detector generates an estimate of the \ac{TX} symbol vector $\mathbf{c}_N$ as
\begin{equation}
\hat{\mathbf{c}}_N \triangleq [\hat{c}_0, \hat{c}_1, ..., \hat{c}_{N-1}]^T = \mathbf{G}_{\mathrm{lmmse}} \, \mathbf{y}_N \text{,}
\label{eq:cN_hat_lmmse}
\end{equation}
where
\begin{equation}
\mathbf{G}_{\mathrm{lmmse}} \triangleq [G^{(\mathrm{lmmse})}_{m,n}] \triangleq \bigl(\mathbf{H}^H \mathbf{H} + \sigma_{\mathrm{w}}^2 \,\mathbf{I}_N \bigr)^{-1} \, \mathbf{H}^H
\label{eq:G_lmmse}
\end{equation}
is the $N \times N$ \ac{LMMSE} equalization matrix. The $n$th element of \eqref{eq:cN_hat_lmmse} can be expressed as
\begin{equation}
\hat{c}_n = \bar{G}_{n,n} \, c_n + \sum_{k\neq n} \bar{G}_{n,k} \, c_k + \bar{w}_n \text{,}
\label{eq:cn_hat_element_lmmse}
\end{equation}
where $\bar{G}_{n,k}$ is the element $(n,k)$ of the $N \times N$ matrix $\bar{\mathbf{G}} \triangleq [\bar{G}_{n,k}] = \mathbf{G}_{\mathrm{lmmse}} \, \mathbf{H}$, with $n$,  $k$ = $0, 1, ..., N-1$.

Note that, in \eqref{eq:cn_hat_element_lmmse}, the term $\bar{w}_n$ is the $n$th element of the vector $\bar{\mathbf{w}}_N \triangleq \mathbf{G}_{\mathrm{lmmse}} \, \mathbf{w}_N$. It is not difficult to show that the \ac{SINR} associated with the estimated symbol $\hat{c}_n$ is $\text{SINR}_n = \bar{G}_{n,n}/(1-\bar{G}_{n,n})$.

Then, the average \ac{BER}, $\text{BER}_{\mathrm{av}}$ for the above mentioned detector can be approximated as
\begin{equation}
\text{BER}_{\mathrm{av}} \cong \frac{1}{N} \frac{4}{\log_2(M_{\mathrm{c}})}\left(1\!-\!\frac{1}{\sqrt{M_{\mathrm{c}}}} \right)\sum_{n=0}^{N-1}Q\left(\sqrt{\frac{3 \, \text{SINR}_n}{M_{\mathrm{c}}\!-\!1}}\right)\! \text{,}
\label{eq:average_ber_lmmse}
\end{equation}
if a \ac{QAM} constellation, with Gray mapping, is assumed for the channel symbols \cite[Sec. 4.3, Eq. (4.3-30)]{proakis_digital_2008};
here, $Q(\cdot)$ denotes the Gaussian Q-function, which can be tightly bounded \cite{AbreuQboundsTC2012}.

\subsection{Cram{\'e}r-Rao Bounds on AFDM Channel Parameter Estimation Errors} \label{subsec:CRB_delay_Doppler}

In this subsection, the \acp{CRB} for the estimation of normalized delay and Doppler in an \ac{AFDM}-based communication system are derived. It is well known that \ac{CRB} provide a fundamental performance benchmark for unbiased estimators; here, they are used to assess the sensing capability of \ac{AFDM} waveforms. The derivation relies on the received signal model in \eqref{eq:yN_vector2} under the following assumptions: 1) The channel consists of a single path\footnote{This assumption leads to closed-form expressions; note that the resulting bounds can be taken as lower bounds for the case of multiple paths.} ($L=1$), characterized by unknown deterministic amplitude $\tilde{a}$, normalized delay $F_{\tau}$ and normalized Doppler $F_{\nu}$; 2) the received signal is affected by \ac{AWGN} with variance $\sigma^2_{\mathrm{w}}$; 3) the vector $\mathbf{c}_N$ contains $N_{\mathrm{u}}\leq N$ non-zero channel symbols, independently drawn from a zero-mean \ac{PSK} or a \ac{QAM} constellation with variance $\sigma^2_{\mathrm{c}}$; 4) the autocorrelation function of $\mathbf{c}_N$ is $\mathbb{E}\{\mathbf{c}_N \,\mathbf{c}_N^H\}
= (N_{\mathrm{u}}/N)\,\sigma^2_{\mathrm{c}} \,\mathbf{I}_N$; 5) a symmetric \ac{FD} pulse and \acp{SC} are employed. 

Under these assumptions, the $2\times 2$ \ac{FIM} associated with the \ac{ML} estimator of $\boldsymbol{\psi}=[F_{\tau},F_{\nu}]^T$ is given by \cite{kay_fundamentals_1993}
\begin{equation}
\mathbf{\Upsilon} \triangleq [\nu_{m,n}]
= 2 \Re \left\{
\frac{\partial \tilde{\mathbf{y}}}{\partial \boldsymbol{\psi}}
\left(\frac{\partial \tilde{\mathbf{y}}}{\partial \boldsymbol{\psi}}\right)^H\right\} \text{,}
\label{eq:FIM}
\end{equation}
where $\tilde{\mathbf{y}} = \mathbf{H} \, \mathbf{c}_N$ denotes the useful component of the received signal. Exploiting the statistical properties of $\mathbf{c}_N$, the $(k,k^{\prime})$th entry of $\mathbf{\Upsilon}$ \eqref{eq:FIM} can be expressed as
\begin{equation}
\nu_{k,k^{\prime}}
= 2 \frac{N{\mathrm{u}}}{N} \sigma^2_{\mathrm{c}}
\, \mathrm{Tr}\left\{
\frac{\partial \mathbf{H}}{\partial \psi_k}
\left(\frac{\partial \mathbf{H}}{\partial \psi_{k'}}\right)^H
\right\} \text{.}
\label{eq:FIM_k_ktilde}
\end{equation}

Closed-form approximations of the considered \acp{CRB} are obtained by: 1) computing the partial derivatives appearing in \eqref{eq:FIM_k_ktilde} for all the combinations of the indices $(k,k^{\prime})$; 2) substituting them into the \ac{FIM} expression; 3) approximating the terms containing a geometric series with the dominant contributions; 4) inverting the resulting \ac{FIM}; 5) extracting its diagonal entries. The detailed derivation and the justification of the adopted approximations are reported in App. \ref{App:CRB_derivation}. After retaining dominant terms, the resulting closed-form approximations of the \acp{CRB} are
\begin{equation}
\text{CRB}_{F_{\tau}} \cong \frac{1}{2\pi^2 \,\sigma^2_{\mathrm{c}}\,\text{SNR} \, \,N_{\mathrm{u}} (\frac{N_{\mathrm{u}}}{3} - 2\lambda_1^2 \, (N-2)^2)}
\label{eq:CRB_delay}
\end{equation}
and
\begin{equation}
\text{CRB}_{F_{\nu}} \cong \frac{N_{\mathrm{u}}}{3\pi^2 \,\sigma^2_{\mathrm{c}}\,\text{SNR} \, N_{\mathrm{u}}(\frac{N_{\mathrm{u}}}{3} - 2\lambda_1^2 \, (N-2)^2)} \text{,}
\label{eq:CRB_Dopp}
\end{equation}

where the \ac{SNR} is defined as $\text{SNR} \triangleq {\vert\tilde{a}\vert^2}/{\sigma^2_{\mathrm{w}}}$. The above expressions lead to the following remarks.

1) If $\lambda_1=0$, the \ac{CRB} refer to \ac{OFDM}.

2) The two expressions are defined and valid for $0\leq \lambda_1 \leq \sqrt{N_{\mathrm{u}}}/(\sqrt{6}(N-2))$. Note that, since $\lambda_1$ is typically selected to be approximately $1/(2N)$ \cite{bemani_affine_2023}, the aforementioned condition is met for $N\geq 16$.

3) In principle, introducing a non-zero chirp parameter $\lambda_1$ in \ac{AFDM} degrades the theoretically achievable accuracy in both delay and Doppler estimation. In fact, when $\lambda_1 = 0$ (i.e., in the \ac{OFDM} case), both \acp{CRB} reduce to their minimum achievable values. In practice, however, in multipath scenarios, conventional estimators are unable to separate different Doppler components in \ac{OFDM}, making the obtained bound unattainable. In contrast, \ac{AFDM} enables separating distinct Doppler components; therefore, in this case, estimators can potentially approach their theoretical performance, despite the higher related \ac{CRB} value.

 
\vspace{-1.5ex}
\section{Numerical Results}\label{Sec:Num_res}

In our work, extensive computer simulations have been run to validate the mathematical results developed in the previous sections. In this section we illustrate various numerical results with the aim of:

1) Showing the impact of pulse-shaping on \ac{AFDM} signal generation; in our analysis different \ac{TX} pulses are taken into consideration and subcarrier suppression is adopted.

2) Assessing the impact of pulse-shaping and of the use of \ac{SC} on the \acp{CRB} derived for normalized delay and Doppler frequency estimation.

3) Comparing our theoretical results about the \ac{BER} performance achieved by \ac{LMMSE} detection with the numerical results obtained when the \ac{DT}-based model available in the literature is adopted. Our comparison concerns two different \ac{DS} channel scenarios and allows us to highlight the impact of fractional delays, pulse-shaping and Doppler modeling accuracy on error performance.

4) Assessing the robustness of the \ac{AFDM} signal processing chain illustrated in Sec. \ref{Sec:Signal_model} in the presence of \ac{HWI}, namely, \ac{PN}, \ac{CFO}, and \ac{SJ}. The \ac{AFDM} error performance is compared with that achieved by \ac{OFDM} in the same conditions. This allows us to quantify the relative resilience of the two modulation schemes in more realistic scenarios.
\vspace{-2ex}
\begin{figure}[H]
\centering
\includegraphics[width=0.9\columnwidth]{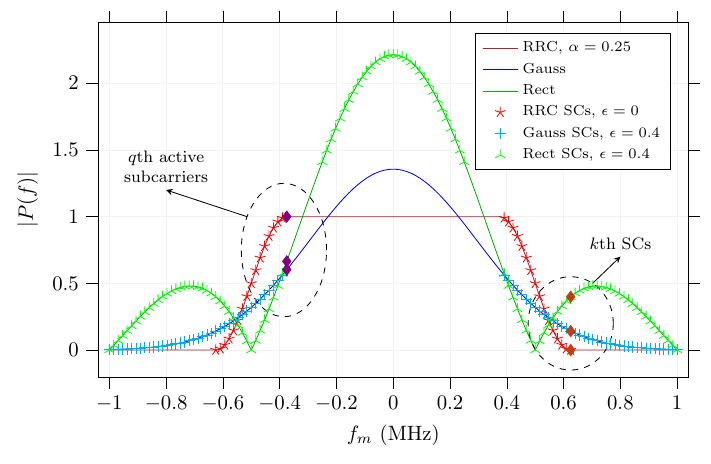}
\vspace{-3ex}
\caption{Magnitude spectrum of the pulses selected for \ac{AFDM} signal generation. The \ac{RRC}, Gaussian and rectangular pulses have been considered. The \ac{AFDM} signal is characterized by a useful bandwidth $B=1$ MHz and $N=64$ subcarriers. Markers indicate the position of the \acp{SC}, so that only $N_{\mathrm{u}}=N-N_{\mathrm{sc}}$ useful subcarriers are retained.}
\label{fig:pulse_choices}
\end{figure}

%
%

\begin{figure*}
\centering
\includegraphics[width=0.9\textwidth]{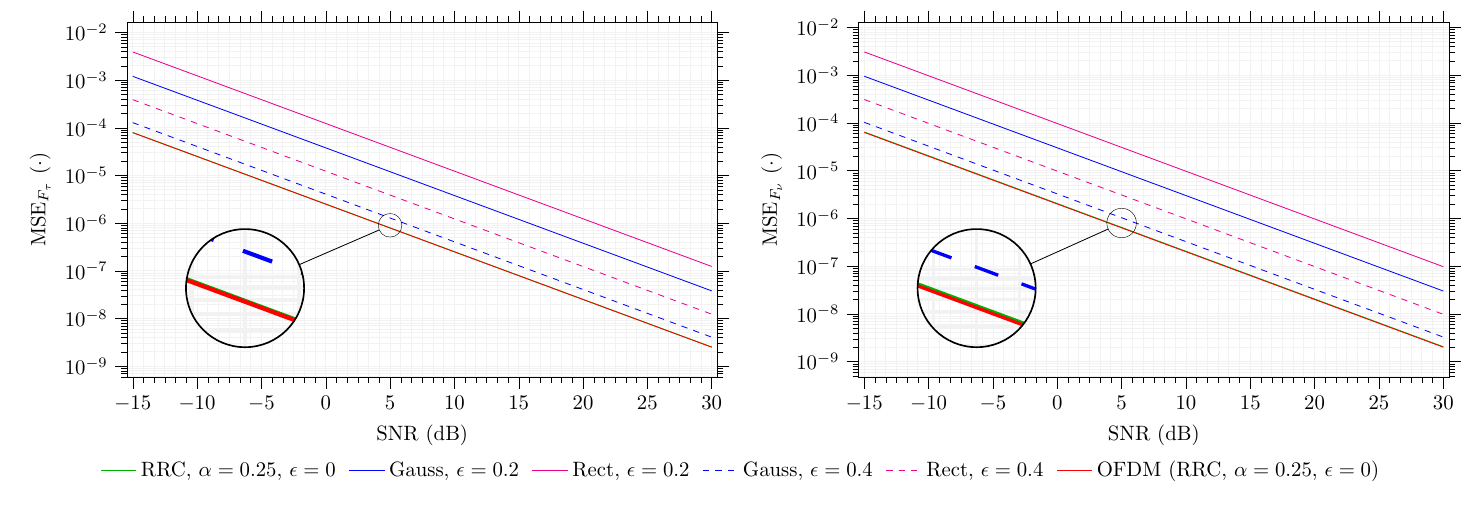}
\vspace{-3ex}
\caption{Theoretical \ac{CRB} versus $\text{SNR} \in [-15, 30]$ for the estimation of normalized delay (left subfigure) and Doppler frequency (right subfigure) in the case of an \ac{AFDM} transmission; three different pulses (namely, \ac{RRC}, Gaussian and rectangular) are used jointly with a \ac{SC} mechanism. The \acp{CRB} achieved by \ac{OFDM} (i.e., for $\lambda_1 =\lambda_2 = 0$) employing \ac{RRC} pulse with $\alpha=0.25$ are also shown for comparison.}
\label{fig:crb_pulses}
\vspace{-3ex}
\end{figure*}

In the remaining part of this section, unless differently stated, the following choices are made: a) an $4$-QAM constellation is employed, so that $\sigma^2_{\mathrm{c}} = 1$; b) the number of subcarriers is $N=64$; c) the subcarrier spacing is $\Delta_f = 15$ kHz; d) the \ac{AFDM} chirp parameter is $\lambda_1 = 0.007$; e) when \ac{RRC} pulse is used, its roll-off factor is $\alpha = 0.25$; f) \ac{CPP} length $N_{\mathrm{cpp}}= 4$; g) carrier frequency $f_{\mathrm{c}}=5.8$ GHz. The values selected for the above mentioned parameters are also listed in Tab. \ref{tab:sim_param_afdm}.

Our first results, shown in Figs. \ref{fig:pulse_choices} and \ref{fig:crb_pulses}, focus on the impact of pulse-shaping on \ac{AFDM} signal generation and on the corresponding estimation-theoretic limits. For such analysis, \ac{RRC}, Gaussian and rectangular pulses, approximately sharing the same bandwidth\footnote{For the rectangular and Gaussian pulses, the bandwidth occupied by the main lobe is considered.} $B \cong 1$ MHz are considered. In particular, the magnitude spectra of the adopted pulses together with the position of the \ac{AFDM} subcarriers are illustrated in Fig. \ref{fig:pulse_choices}. Is important to note that:

1) Only the \ac{RRC} pulse features a flat-top frequency response over its useful band, ensuring equal amplitude across all active subcarriers; this property is not shared by the other two pulses.

2) The \ac{SC} mechanism explained in Sec. \ref{subsec:tx_afdm_signal} is applied in order to prevent \ac{SI}. For \ac{RRC} pulses, the number of \acp{SC} is determined by the roll-off factor, leading to $N_{\mathrm{u}} = 2N_{\alpha}+1$ useful subcarriers.
In turn, for Gaussian and rectangular pulses, the absence of a flat-top region requires introducing a positive tolerance parameter, here fixed to $\epsilon = 0.4$, in order to establish which subcarriers can be retained. As highlighted in Fig. \ref{fig:pulse_choices}, the subcarrier frequencies are indexed by ${m=q+kN}$, consistently with \eqref{eq:AFDM_pulse_expanded}. This representation clearly shows that, when shifting the $q$th active subcarrier (see the purple diamonds in Fig. \ref{fig:pulse_choices}) by multiples of $N$, either the corresponding spectral replica must be suppressed or the pulse spectrum must be null at the corresponding frequency (see, for instance, the brown diamonds, referring to \acp{SC}, in Fig. \ref{fig:pulse_choices}). Otherwise, more than a single term survives in the summation over $k$ in \eqref{eq:AFDM_pulse_expanded}, so that the resulting waveform does not exhibit a multicarrier structure.
Preserving the multicarrier structure of \ac{AFDM} requires a strictly band-limited pulse and an \ac{SC} mechanism; among the examined options, the \ac{RRC} pulse is the most suitable thanks to its flat-top region.

In Fig. \ref{fig:crb_pulses} the \acp{CRB} for the estimation of the normalized delay and the Doppler frequency (see \eqref{eq:CRB_delay} and \eqref{eq:CRB_Dopp}, respectively) are illustrated for each of the three considered pulse shapes; the \acp{CRB} for \ac{OFDM} are also shown for comparison. Note that our derivation of \ac{CRB} relies \ac{AWGN} with identical statistics across subcarriers. When non-flat pulses are used, the retained subcarriers are scaled differently, thus inducing non-uniform noise statistics across samples. Hence, the \ac{AWGN} assumption underlying the \ac{CRB} no longer holds. Therefore, for non-flat pulses, the derived expressions should be regarded as optimistic lower bounds, since the effective noise becomes non-uniform across subcarriers and the exact \ac{FIM} would require accounting for the resulting colored-noise covariance. Based on these results, the following comments can be made:

1) In the case of \ac{AFDM}, the \ac{RRC} pulse consistently achieves the lowest \ac{CRB} values in both delay and Doppler estimation over the entire \ac{SNR} range. This is a direct consequence of its flat-top frequency response, which ensures uniform subcarrier weighting and maximizes the useful number of samples $N_{\mathrm{u}}$.

2) For the same tolerance $\epsilon$, the Gaussian pulse outperforms the rectangular one. The smoother spectral decay of the Gaussian pulse provides a more favorable distribution of spectral energy across the retained subcarriers, resulting in improved estimation accuracy.

3) Increasing the tolerance $\epsilon$ enlarges the number of useful subcarriers $N_{\mathrm{u}}$, thereby increasing the number of signal samples contributing to delay and Doppler estimation.

4) The \acp{CRB} obtained for \ac{OFDM} are very close to those of \ac{AFDM} for the same \ac{RRC} pulse. Keep in mind, however, that this comparison refers to the case of single-path (i.e., $L=1$); in multipath scenarios, \ac{OFDM}, unlike \ac{AFDM}, is unable to separate the contributions of different paths.

\begin{table}
\centering
\caption{Parameters selected for \ac{AFDM} in our computer simulations.}
\vspace{-1ex}
\label{tab:sim_param_afdm}
\begin{tabular}{ll}
\hline
\textbf{Parameter} & \textbf{Value} \\
\hline
$\#$ subcarriers & $N = 64$ \\
\ac{AFDM} chirp parameters & $\lambda_1 = \lambda_2 = 0.007$ \\
\ac{CPP} length & $N_{\mathrm{cpp}} = 4$ \\
Pulse-shaping & \ac{RRC}, with $\alpha = 0.25$ \\
Constellation & $4$-QAM \\
Carrier frequency & $f_{\mathrm{c}} = 5.8$ GHz \\
Subcarrier spacing & $\Delta_f = 15$ kHz \\
\hline
\end{tabular}
\end{table}

In conclusion, the \ac{RRC} pulse emerges as the most suitable choice among the considered candidates. In general, pulses exhibiting a flat-top magnitude spectrum are preferable for \ac{AFDM} modulation, as they simultaneously guarantee \ac{SI} mitigation and uniform subcarrier weighting, thus leading to improved performance in parameter estimation.


Let us compare now the \ac{BER} performance obtained in the case of the developed \ac{CT} channel model with that derived from the \ac{DT} formulation commonly adopted in the literature. In this case, \ac{AFDM} signals are generated using only the \ac{RRC} pulse and the wireless channel is assumed to be \ac{DS}: its Doppler spectrum is generated according to the Jakes model, with maximum Doppler spread determined by the maximum channel path velocity $v_{\mathrm{max}}$. The \ac{PDP} and the associated path delays follow the \emph{tapped delay line} model of \emph{type A} (TDL-A) \cite{3gpp_study_2022}.

\begin{figure}[H]
\centering
\includegraphics[width=0.9\columnwidth]{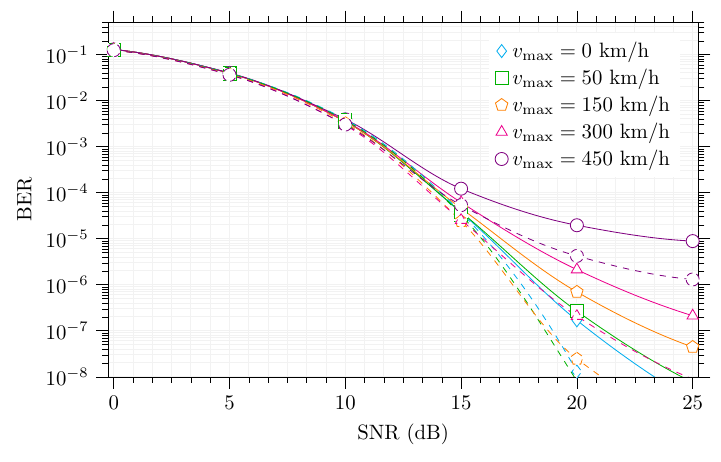}
\vspace{-3ex}
\caption{Bit-error rate achieved by a \ac{LMMSE} detector for \ac{AFDM} in a \ac{DS} channel with $L=3$ paths and maximum velocity $v_{\mathrm{max}} \in \{0, 50, 150, 300, 450\}$ km/h. Solid lines correspond to the channel matrix obtained from the proposed \ac{CT} model, while dashed lines refer to its \ac{DT} model-based counterpart.}
\label{fig:ber_snr_vmax}
\vspace{1ex}
\includegraphics[width=0.9\columnwidth]{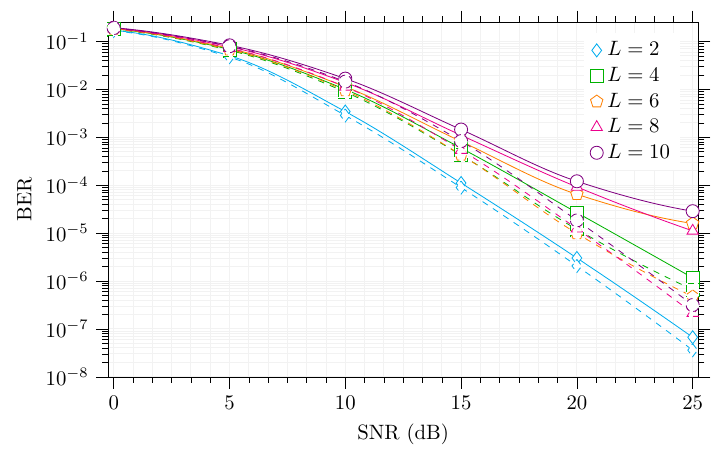}
\vspace{-3ex}
\caption{Bit-error rate achieved by a \ac{LMMSE} detector for \ac{AFDM} in the presence of a \ac{DS} channel with number of paths $L\in \{2,4,6,8,10\}$. Solid lines correspond to the channel matrix obtained from the proposed \ac{CT} model, while dashed lines refer to its \ac{DT} model-based counterpart.}
\label{fig:ber_snr_L}
\vspace{-2ex}
\end{figure}

The \ac{BER} curves\footnote{Formula \eqref{eq:average_ber_lmmse} has been exploited to generate these results; averaging over $100$ independent Monte Carlo channel realizations has been accomplished.} obtained for \ac{LMMSE} detection in the presence of the maximum channel path velocity  $v_{\mathrm{max}}\in\{0,50,150,300,450\}$ km/h are shown in Fig. \ref{fig:ber_snr_vmax}. In this case, both the channel matrix derived from the proposed \ac{CT} model (solid lines) and that obtained from the \ac{DT} model \cite[Eq. (30)]{bemani_affine_2023} (dashed lines) are shown.

These results show that the \ac{BER} performance degrades as $v_{\mathrm{max}}$ increases because of the larger Doppler spread, which results in a higher time selectivity and stronger \ac{ICI}. More importantly, the \ac{BER} values predicted on the basis of the \ac{CT} model are consistently higher than those obtained with the \ac{DT} model. The resulting gap highlights a fundamental limitation of \ac{DT}-based models, which neglect both the impact of pulse-shaping and the presence of fractional delays (the presence of fractional Doppler shifts only is considered).


Other \ac{BER} results are shown in Fig. \ref{fig:ber_snr_L}, which unveils the impact of the overall number of paths\footnote{As $L$ increases, the overall delay spread generally enlarges according to the adopted TDL-A profile, leading to increased frequency selectivity.} $L$ on \ac{AFDM} performance; in this case, $L\in\{2,4,6,8,10\}$ and $v_{\mathrm{max}}=250$ km/h. It is easily inferred that increasing the number of channel paths, $L$, results in a progressive \ac{BER} degradation. A larger $L$ implies a richer multipath structure and enhanced \ac{DD} dispersion, thus worsening detection performance. Moreover, once again, the \ac{BER} values obtained with the more accurate \ac{CT} model are consistently higher than those achieved with the \ac{DT} model for every considered value of $L$ across the entire \ac{SNR} range. This confirms that the \ac{DT} formulation systematically underestimates the impact of the multipath components of a \ac{DS} channel.

\begin{figure}[H]
\centering
\includegraphics[width=0.9\columnwidth]{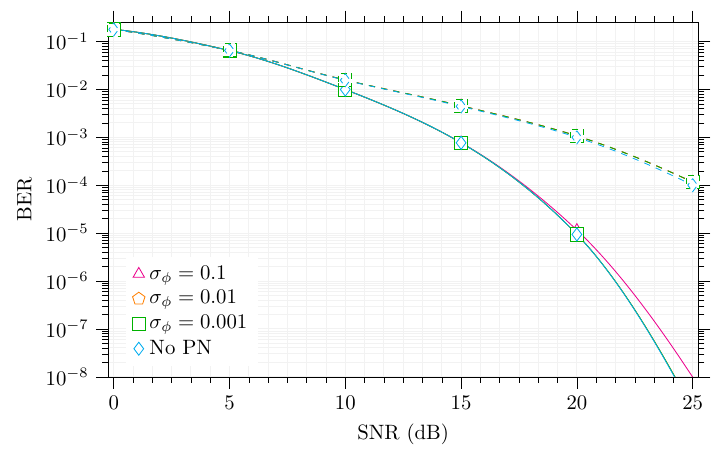}
\vspace{-3ex}
\caption{Average \ac{BER} observed in the presence of \ac{PN} for \ac{AFDM} (solid lines) and \ac{OFDM} (dashed lines); $\sigma_{\phi} \in \{0, 0.001, 0.01, 0.1\}$ is considered.}
\label{fig:ber_pn}
%
\vspace{1ex}
\includegraphics[width=0.9\columnwidth]{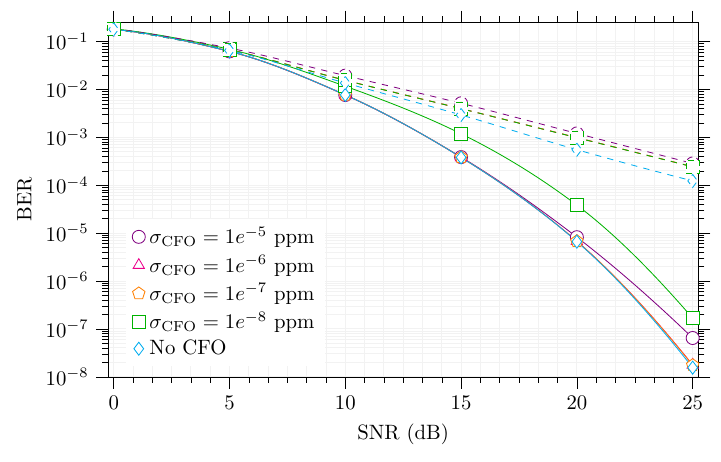}
\vspace{-3ex}
\caption{Average \ac{BER} observed in the presence of \ac{CFO} for \ac{AFDM} (solid lines) and \ac{OFDM} (dashed lines). The \ac{CFO} \acp{std} are expressed in ppm relative to the carrier frequency $f_{\mathrm{c}}$.}
\label{fig:ber_cfo}
%
\vspace{1ex}
\includegraphics[width=0.9\columnwidth]{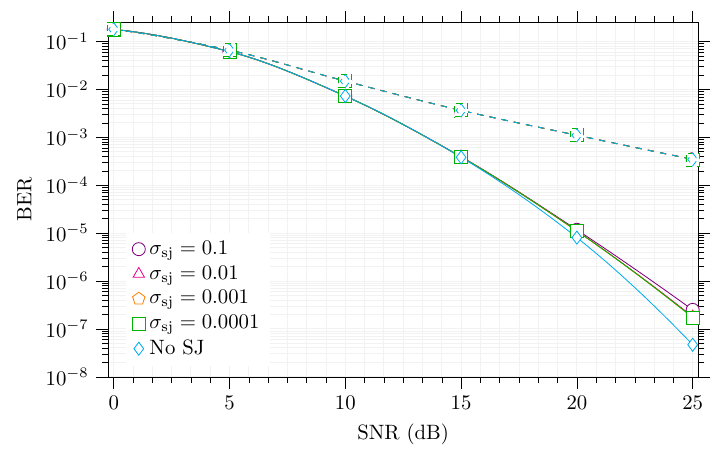}
\vspace{-3ex}
\caption{Average \ac{BER} observed in the presence of \ac{SJ} for \ac{AFDM} (solid lines) and \ac{OFDM} (dashed lines); $\sigma_{\mathrm{sj}} \in \{0.1, 0.01, 0.001, 0.0001\}$ is considered.}
\label{fig:ber_sj}
\end{figure}

Our final results, illustrated in Figs. \ref{fig:ber_pn}-\ref{fig:ber_sj}, allow us to compare the \ac{AFDM} error performance with the \ac{OFDM} one in the presence of \ac{HWI}, namely \ac{PN}, \ac{CFO}, and \ac{SJ}. To ensure a fair comparison, the same \ac{LMMSE} detector has been used for both modulation formats. Moreover, the following choices have been made: a) we have selected a \ac{DS} channel, whose \ac{PDP} follows a TDL-A channel model with $L=3$ paths, a delay spread of $0.5$ \textmu s, and a maximum velocity $v_{\mathrm{max}}=250$ km/h; b) zero-mean Gaussian distributions with \acp{std} $\sigma_{\phi}$, $\sigma_{\mathrm{CFO}}$, and $\sigma_{\mathrm{sj}}$, respectively, have been employed to model the impairment processes.

The impact of \ac{PN} on the system performance is exemplified by Fig. \ref{fig:ber_pn}. From these results, it can be easily inferred that \ac{AFDM} significantly outperforms \ac{OFDM} for all the considered values of $\sigma_{\phi}$. While \ac{OFDM} suffers from a severe error floor, even at low \ac{PN} levels, due to both common phase error and Doppler-induced \ac{ICI}, \ac{AFDM} maintains a performance close to the case of no \ac{PN} for $\sigma_{\phi} \leq 0.01$, thus demonstrating its superior robustness to phase instabilities.

The impact of \ac{CFO} on error performance is evidenced by Fig. \ref{fig:ber_cfo}, which the \ac{BER} curves obtained for different values $\sigma_{\mathrm{CFO}}$. These results show that \ac{OFDM} performance degrades rapidly as the \ac{CFO} increases and, in particular, that the \ac{BER} curves exhibit a floor higher than $10^{-3}$ for $\sigma_{\text{CFO}} \geq 10^{-7}$ ppm. This does not occur in the case of \ac{AFDM}. In fact, even in the worst case (i.e., for $\sigma_{\mathrm{CFO}}=10^{-5}$ ppm), a \ac{BER} close to $10^{-7}$ is observed for an $\text{SNR}=25$ dB; on the contrary, \ac{OFDM} is unable to provide reliable communication in the same \ac{SNR} range.

Our last results, illustrated in Fig. \ref{fig:ber_sj}, concern the impact of \ac{SJ} on error performance. They show that both modulations are relatively robust to the variability of the \ac{SJ} itself, as the curves within each group (solid or dashed) remain closely bundled. The substantially higher \ac{BER} of \ac{OFDM} compared to \ac{AFDM} is mainly due to the Doppler-induced \ac{ICI} originating from the \ac{DS} channel, rather than from the \ac{SJ} itself. Interestingly, \ac{AFDM} appears more sensitive to \ac{SJ} variations than \ac{OFDM}, as evidenced by the wider spread observed in its \ac{BER} curves at high \ac{SNR} values (above $20$ dB). This increased sensitivity originates from the chirp-based nature of the \ac{AFDM} waveform.

\vspace{-4ex}
\section{Conclusions}
\label{Sec:Concl}

In this manuscript, a comprehensive \ac{CT} analytical framework for \ac{AFDM} modulation has been developed. Based on it, a number of new results have been obtained.
First of all, our analysis has evidenced that carefully designed pulse-shaping and subcarrier suppression are needed to preserve the multicarrier structure of \ac{AFDM} signals.
Secondly, the \ac{CT}-based analysis of such signals has allowed us derive a closed form expression for the \ac{PSD} of \ac{AFDM}. Based on it, we have shown that, although \ac{AFDM} shares some spectral similarities with \ac{OFDM}, pulse-shaping and its truncation play a critical role, since they significantly influence \ac{OOB}, due to the additional chirp modulation inherent to \ac{AFDM}.
Thirdly, we have shed new light on the impact of \acp{HWI} (including \ac{PN}, \ac{CFO}, and \ac{SJ}) on \ac{AFDM}. Our result evidence that, although not impervious to these effects, \ac{AFDM} modulation format maintains a significant resilience compared to conventional \ac{OFDM} in high-mobility channels; this confirms what the technical literature claims on \ac{AFDM}.
Fourthly, closed-form expressions of the \acp{CRB} for delay and Doppler estimation have been derived for the case of a single path channel. They highlight a fundamental trade-off.
In fact, although \ac{AFDM} exhibits a higher theoretical estimation variance than \ac{OFDM} due to its additional chirp modulation, it enables the resolution of Doppler ambiguities in multipath environments.
These results provide useful tools for the design and realistic performance assessment of next-generation wireless communication systems.

Future work will focus on developing channel estimation and data detection techniques based on the proposed \ac{CT} model.
Such techniques are expected to enable a more efficient design of pilot signaling and the development of new \ac{AFDM} receiver architectures, thus providing new insights into \ac{AFDM} feasibility over state-of-the-art implementations.

\vspace{-1ex}
\begin{appendix}
\section{Appendix}

\subsection{Properties of AFS and Continuous AFT}
\label{APP:AFS_AFT}

In this appendix, we summarize the main properties and characteristics of the \ac{AFS} and the continuous \ac{AFT}, which generalize classical Fourier transforms for chirp-periodic signals.

Let $s(t)$ be a chirp-periodic signal with chirp period $T$ and chirp parameter $\bar{\lambda}_1$, such that, for the given $T$, $s(t+T) = s(t)\, \exp(j 2 \pi \bar{\lambda}_1 (T^2 + 2 T t))$, holds for any $t$.

Then, the \ac{AFS} expansion of $s(t)$ is $s(t) = \sum_{m=-\infty}^{+\infty} S_m \, \psi_m(t)$, where the $m$th basis functions is
\begin{equation}
\psi_m(t) = \exp\left(j 2 \pi \left(\bar{\lambda}_1 t^2 + \frac{m}{T} t\right)\right)
\label{APP:basis_func_psi_mt}
\end{equation}
for any $m$ and the $m$th coefficient is evaluated as
\begin{equation}
S_m = \frac{1}{T} \int_0^T s(t) \, \psi_m^{\ast}(t) \, {\rm d}t\text{.}
\end{equation}

Note that the \ac{AFS} reduces to the classical \ac{FS} when $\bar{\lambda}_1=0$. Moreover, the basis functions $\psi_m(t)$, in \eqref{APP:basis_func_psi_mt}, are orthogonal over one chirp period, i.e., $\int_0^T \psi_m(t) \, \psi_n^{\ast}(t) \, {\rm d}t= 0$, for $ m\neq n$.

The continuous \ac{AFT} generalizes the \ac{CFT} to signals with quadratic phase.
The operation, dependent on the parameter $\bar{\lambda}_1$, can be written as
\begin{equation}
S_{\lambda_1}(f) = \int_{-\infty}^{+\infty} s(t) \, \exp\left(-j 2 \pi \left(\bar{\lambda}_1 t^2 + f t\right)\right) \, {\rm d}t\text{.}
\label{eq:AFT_formula}
\end{equation}

This last operation maps a chirp-modulated \ac{TD} signal into a domain where \ac{DD} characteristics are separated. Moreover, it reduces to the classical \ac{CFT} when $\bar{\lambda}_1=0$. It is also worth noting that the \ac{AFT} of the sum of signals is the sum of their \acp{AFT} (i.e., the \emph{linearity} property holds). Finally, the inversion formula is
\begin{equation}
s(t) = \int_{-\infty}^{+\infty} S_{\lambda_1}(f) \, \exp\left(j 2 \pi \left(\bar{\lambda}_1 t^2 + f t\right)\right) \, {\rm d}f \text{.}
\end{equation}

\subsection{Filtering of a Chirp-Exponential Waveform} \label{APP_convolution_with_chirp}

In this appendix we concentrate on the response of a filter matched to the signal $p(t)$ to the chirp-exponential function

\begin{equation}
u(t) \triangleq \exp\Bigl(j2\pi\Bigl(\bar{\lambda}_1 t^2 + \frac{m}{T}t\Bigr)\Bigr) \text{.}
\label{eq:u_t}
\end{equation}

In particular, we consider the linear convolution
\begin{equation}
y(t) = p^{\ast}(-t) \ast u(t) = \int_{-\infty}^{+\infty} p(-\tau)\, u(t-\tau)\, {\rm d}\tau \text{.}
\label{eq:conv_with_chirp_exp}
\end{equation}
Substituting the \ac{RHS} of \eqref{eq:u_t} in that of the last formula and expanding the quadratic phase yields
\begin{eqnarray}
y(t)\!=&&\hspace{-5ex} \int_{-\infty}^{+\infty}\!\!\!\!\!\! p(-\tau) \exp\Bigl(j2\pi\Bigl(\bar{\lambda}_1 (t-\tau)^2 + \frac{m}{T}(t-\tau)\Bigr)\Bigr) \, {\rm d}\tau \\
=&&\hspace{-3ex} \exp\Bigl(j2\pi\Bigl(\bar{\lambda}_1 t^2 + \frac{m}{T}t\Bigr)\Bigr)\nonumber \\
&&\cdot \int_{-\infty}^{+\infty}\!\!\!\!\!\! p(-\tau)\exp\Bigl(j2\pi\Bigl(\bar{\lambda}_1 \tau^2 -2\bar{\lambda}_1 t\,\tau - \frac{m}{T}\tau\Bigr)\Bigr) \, {\rm d}\tau \text{.}\nonumber
\end{eqnarray}

Changing the integration variable in the last integral by defining $\sigma =-\tau$ gives
\begin{eqnarray}
y(t) = \exp\Bigl(j2\pi\Bigl(\bar{\lambda}_1 t^2 + \frac{m}{T}t\Bigr)\Bigr)&& \\
&&\hspace{-25ex}\cdot \int_{-\infty}^{+\infty} p(\sigma)\,\exp\Bigl(j2\pi\Bigl(\bar{\lambda}_1 \sigma^2 +\Bigl(2\bar{\lambda}_1 t + \frac{m}{T}\Bigr)\sigma\Bigr)\Bigr) \, {\rm d}\tau \text{.}
\nonumber
\end{eqnarray}

Let now define $f_{m}(t) \triangleq 2\bar{\lambda}_1 t + m/T$, that represents $m$th time varying frequency. Then, keeping in mind the definition of \ac{AFT} (see \eqref{eq:AFT_formula}), we can write that $y(t) = u(t) \, P_{\lambda_1}^{\ast}(f_{m}(t))$. This proves that the convolution with the chirp-exponential defined in \eqref{eq:conv_with_chirp_exp} cannot be seen as a time-invariant filtering in the usual sense, since the output is the chirp-exponential $u(t)$ multiplied by the \ac{AFT} of $p(t)$ evaluated along the \emph{time-varying} frequency argument $f_{m}(t)$.

\vspace{-2ex}
\subsection{Impulse Response Characterization of an AFDM DS Channel}
\label{App:AFD_IR}

To provide further insight into the structure of the effective \ac{AFDM} channel matrix $\mathbf{H} \triangleq [H_{m,n}]$ (see \eqref{eq:Hmn_doubly_selective1}) derived in Sec. \ref{subsec:rx_afdm_signal_ds}, we adopt an \ac{IR}-based representation in the \ac{AFD}. In particular, we characterize the effective channel by evaluating its response to a discrete Dirac delta, i.e.,
\begin{equation}
\mathbf{h}^{(\mathrm{IR})} \triangleq \Bigl[h^{(\mathrm{IR})}_0, h^{(\mathrm{IR})}_1, ..., h^{(\mathrm{IR})}_{N-1}\Bigr]^T = \mathbf{H} \, \boldsymbol{\delta}_N,
\label{eq:AFD_IR_def}
\end{equation}
where $\boldsymbol{\delta}_N$ is the $N$-dimensional Dirac vector (its entries are all zero except for a single unitary element\footnote{Without loss of generality, the index of non-zero element corresponds to the center of the vector, i.e., is $n = \lfloor N/2\rfloor$.}.

Note that the vector $\mathbf{h}^{(\mathrm{IR})}$, in \eqref{eq:AFD_IR_def}, provides insights on how the energy of an isolated \ac{AFDM} impulse spreads across that domain due to the combined effect of the \ac{DS} channel and the \ac{AFDM} modulation. In fact, it allows us to assess: 1) the path localization properties of the effective channel in the \ac{AFD}; 2) the Doppler-induced \ac{ICI} profile; 3) the impact of the chirp parameter $\lambda_1$ on the dispersion of the \ac{CIR}.

\begin{figure}[H]
\centering
\includegraphics[width=0.9\columnwidth]{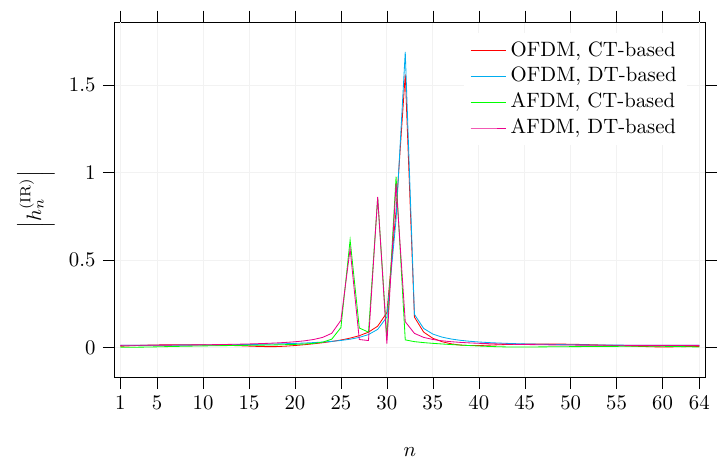}
\vspace{-3ex}
\caption{Magnitude of the \ac{AFDM} \ac{IR} $\vert h^{(\mathrm{IR})}_n \vert$ for the \ac{OFDM} and \ac{AFDM} channel matrices obtained from \ac{CT}-based and \ac{DT}-based signal models.}
\label{fig:IR_ds_3paths}
\vspace{-1ex}
\end{figure}

Moreover, the same \ac{IR}-based characterization can be easily extended to \ac{AFDM} channel matrices that include receiver \acp{HWI}. In this case, the corresponding \ac{AFDM} \ac{IR} provides an insight on how such impairments affect energy spreading and coupling in the \ac{AFD}.

The magnitude of the \acp{IR} $\mathbf{h}^{(\mathrm{IR})}$ obtained from both the proposed \ac{CT}-based model and the \ac{DT}-based \ac{AFDM} formulation are illustrated in Fig. \ref{fig:IR_ds_3paths}; for both \ac{AFDM} and \ac{OFDM} (corresponding to $\lambda_1 = \lambda_2 = 0$). In this case, a single realization of a \ac{DS} channel, characterized by $L=3$ propagation paths with delay taps $\tau_{\mathrm{bin}}$ equal to $1.1$ (including fractional delays), $3$ and $6$, and a Doppler spectrum following a Jakes model with maximum Doppler shift $\nu_{\mathrm{max}}=12$ kHz\footnote{According to the parameters listed in Tab. \ref{tab:sim_param_afdm}, this value corresponds to a maximum path velocity $v_{\mathrm{max}}\cong 1120$ km/h.} is considered. Moreover, the \ac{AFDM} parameters employed to generate these numerical results are summarized in Tab. \ref{tab:sim_param_afdm}.

From Fig. \ref{fig:IR_ds_3paths}, the following conclusions can be inferred: 1) the close match between \ac{CT}-based and \ac{DT}-based \acp{IR} for both \ac{AFDM} and \ac{OFDM} confirms the consistency of the \ac{CT} formulation adopted in this work; 2) compared to \ac{OFDM}, \ac{AFDM} exhibits a more structured spreading of energy across \ac{AFD} indices, induced by the chirp modulation and captured by the Dirichlet kernel terms in the effective channel matrix; 3) the curves associated with \ac{OFDM} show the \acp{IR} in the usual \ac{FD}, where the contribution of each of the $3$ channel paths cannot be easily separated.


\subsection{Derivation of the Cram{\'e}r-Rao Bounds} \label{App:CRB_derivation}
In this appendix provides the derivation of the closed-form approximations of the \acp{CRB} provided in Subsec. \ref{subsec:CRB_delay_Doppler} is sketched.

We refer to the same noisy signal model \eqref{eq:yN_vector2} and parameter vector $\boldsymbol{\psi}=[F_{\tau},F_{\nu}]^T$ adopted in the aforementioned subsection. The entry $(m,n)$ of the channel matrix $\mathbf{H}$ is given in \eqref{eq:Hmn_doubly_selective1} and can be written in compact form as
\begin{equation}
H_{m,n}
= \frac{1}{N} B_{m,n}\,\tilde{a}\,
\exp(j2\pi m F_{\tau})\,
S_{m,n}(F_{\tau},F_{\nu}) \text{,}
\end{equation}
where
\begin{equation}
S_{m,n}(F_{\tau},F_{\nu}) = \frac{1-r^N}{1-r}
\end{equation}
and
\begin{equation}
r = \exp\!\left(-j2\pi\!\left(\frac{F_{\nu}}{N}
+2\lambda_1 NF_{\tau}
-\frac{n-m}{N}\right)\right) \text{.}
\label{eq:r_def_geom_ser}
\end{equation}

The derivatives of $H_{m,n}$ with respect to the normalized delay and Doppler are
\begin{equation}
\frac{\partial H_{m,n}}{\partial F_{\tau}}
= B_{m,n}\frac{j2\pi}{N}\tilde{a}
\exp(j2\pi mF_{\tau})
\left(m - 2\lambda_1 T_{m,n}\right) \text{,}
\label{eq_der_1}
\end{equation}
and
\begin{equation}
\frac{\partial H_{m,n}}{\partial F_{\nu}}
= -B_{m,n} \, \frac{j2\pi}{N} \, \tilde{a}
\exp\bigl(j2\pi mF_{\tau}\bigr)\,
T_{m,n} \text{,}
\label{eq_der_2}
\end{equation}
respectively; here,
\begin{equation}
T_{m,n}
= \frac{r}{(1-r)^2}
\left(1-N \, r^{N-1}+(N-1)\, r^N\right) \text{.}
\end{equation}

Note that, when the phase term in \eqref{eq:r_def_geom_ser} approaches zero for a given $(m,n)$ (i.e., $r\rightarrow 1$), the geometric series terms admit the limits
\begin{equation}
S_{m,n} \rightarrow N \text{,}
\qquad
T_{m,n} \rightarrow \frac{N(N-1)}{2} \text{.}
\end{equation}

Let us now substitute the \ac{RHS} of \eqref{eq_der_1} and \eqref{eq_der_2} in that of \eqref{eq:FIM_k_ktilde}. Then, it is found that the entries of the $2\times2$ \ac{FIM} involve double sums over $(m,n)$; moreover, such sums contain terms in $|m-2\lambda_1 T_{m,n}|^2$, $|T_{m,n}|^2$ and their cross-products.  

Under the assumptions of \ac{FD} symmetry for the adopted pulse and use of \acp{SC} (as already mentioned in Subsec. \ref{subsec:CRB_delay_Doppler}), $|B_{m,n}|^2$ can be deemed approximately constant over the $N_{\mathrm{u}}$ active subcarriers. Moreover, the dominant contribution to the above mentioned sums originates from the index pairs for which the phase in \eqref{eq:r_def_geom_ser} is small (since these are associated with the constructive contributions of the geometric series); in fact, the remaining terms yield negligible contribution for moderate-to-large $N$ (namely, $N\geq 16$). For these dominant terms, we adopt the approximation $T_{m,n} \approx N(N-1)/2$, so that the leading contribution to the $(0,0)$ \ac{FIM} entry is proportional to
\begin{equation}
\sum_{m=0}^{N_{\mathrm{u}}-1} m^2
= \frac{N_{\mathrm{u}}(N_{\mathrm{u}}-1)(2N_{\mathrm{u}}-1)}{6}
\approx \frac{N_{\mathrm{u}}^3}{3} \text{,}
\end{equation}
if we retain dominant polynomial terms only. Performing similar simplifications for all the remaining \ac{FIM} entries leads, after computing the inverse of the resulting \ac{FIM}, to the closed-form \ac{CRB} expressions \eqref{eq:CRB_delay} and \eqref{eq:CRB_Dopp} for the normalized delay and Doppler, respectively.

It is worth mentioning that, when $\lambda_1=0$, the \ac{AFDM} model reduces to \ac{OFDM}. In this case, the cross-coupling terms vanish and the \ac{FIM} becomes diagonal; this yields
\begin{equation}
\text{CRB}_{F_{\tau}}^{(\mathrm{OFDM})}
= \frac{3}{2\pi^2 \, \sigma_c^2 \,\mathrm{SNR} \, N_{\mathrm{u}}^2}
\label{App:CRB_ftau_OFDM}
\end{equation}
and
\begin{equation}
\text{CRB}_{F_{\nu}}^{(\mathrm{OFDM})}
= \frac{1}{2\pi^2 \, \sigma_c^2 \, \mathrm{SNR}\, N_{\mathrm{u}}} \text{,}
\label{App:CRB_fnu_OFDM}
\end{equation}
which coincide with the expressions obtained by setting $\lambda_1=0$ in the \ac{AFDM} \acp{CRB}, with $\text{SNR} \triangleq {\vert\tilde{a}\vert^2}/{\sigma^2_{\mathrm{w}}}$.

\end{appendix}


%
%


\end{document}

%% file: acro_list.tex
\newacronym{OFDM}{OFDM}{orthogonal frequency division multiplexing}
\newacronym{DS}{DS}{doubly-selective}
\newacronym{DFT-s-OFDM}{DFT-s-OFDM}{DFT-spread \ac{OFDM}}
\newacronym{FBMC}{FBMC}{filter bank multi-carrier}
\newacronym{GFDM}{GFDM}{generalized frequency division multiplexing}
\newacronym{OCDM}{OCDM}{orthogonal chirp division multiplexing}
\newacronym{AFDM}{AFDM}{affine frequency division multiplexing}
\newacronym{OTFS}{OTFS}{orthogonal time frequency space}
\newacronym{T-OTFS}{T-OTFS}{transcendentally-rotated OTFS}
\newacronym{ODDM}{ODDM}{orthogonal delay-Doppler division multiplexing}
\newacronym{SWHM}{SWHM}{sparse Walsh-Hadamard multiplexing}
\newacronym{FMCW}{FMCW}{frequency modulated continuous wave}
\newacronym{OTSM}{OTSM}{orthogonal time sequency modulation}

\newacronym{LCT}{LCT}{linear canonical transform}
\newacronym{ZT}{ZT}{Zak transform}
\newacronym{DZT}{DZT}{discrete Zak transform}
\newacronym{IDZT}{IDZT}{inverse discrete Zak transform}
\newacronym{FT}{FT}{Fourier transform}
\newacronym{IFT}{IFT}{inverse Fourier transform}
\newacronym{DFT}{DFT}{discrete Fourier transform}
\newacronym{IDFT}{IDFT}{inverse discrete Fourier transform}
\newacronym{AFT}{AFT}{affine Fourier transform}
\newacronym{DAFT}{DAFT}{discrete affine Fourier transform}
\newacronym{IDAFT}{IDAFT}{inverse discrete affine Fourier transform}
\newacronym{SFFT}{SFFT}{symplectic finite Fourier transform}
\newacronym{ISFFT}{ISFFT}{inverse symplectic finite Fourier transform}
\newacronym{HT}{HT}{Heisenberg transform}
\newacronym{WT}{WT}{Wigner transform}
\newacronym{frFT}{frFT}{fractional Fourier transform}
\newacronym{IfrFT}{IfrFT}{inverse fractional Fourier transform}
\newacronym{fnT}{fnT}{Fresnel transform}
\newacronym{IfnT}{IfnT}{inverse Fresnel transform}
\newacronym{LT}{LT}{Laplace transform}
\newacronym{ILT}{ILT}{inverse Laplace transform}

\newacronym{FP}{FP}{Fractional Programming}
\newacronym{LDT}{LDT}{Lagrangian Dual Transform}
\newacronym{QT}{QT}{Quadratic Transform}

\newacronym{TDL-A}{TDL-A}{tapped delay line model of type A}
\newacronym{ISAC}{ISAC}{integrated sensing and communications}
\newacronym{JCAS}{JCAS}{joint communications and sensing}
\newacronym{EM}{EM}{electromagnetic}
\newacronym{CP}{CP}{cyclic prefix}
\newacronym{B5G}{B5G}{beyond fifth generation}
\newacronym{3GPP}{3GPP}{$3^{\rm{rd}}$ generation partnership project} 
\newacronym{4G}{4G}{fourth generation}                                
\newacronym{5G}{5G}{fifth generation}                                 
\newacronym{6G}{6G}{sixth generation}
\newacronym{LTV}{LTV}{linear time-variant}
\newacronym{LTI}{LTI}{linear time-invariant}
\newacronym{LTVM}{LTVM}{linear time-variant multipath}
\newacronym{TV}{TV}{time-variant}
\newacronym{TI}{TI}{time-invariant}
\newacronym{1D}{1D}{one-dimensional}
\newacronym{2D}{2D}{two-dimensional}
\newacronym{3D}{3D}{three-dimensional}
\newacronym{NTN}{NTN}{non-terrestrial network}
\newacronym{LEO}{LEO}{low earth orbit}
\newacronym{IoT}{IoT}{Internet-of-Things}
\newacronym{mmWave}{mmWave}{millimeter-wave}
\newacronym{THz}{THz}{Terahertz}
\newacronym{V2X}{V2X}{vehicle-to-everything}
\newacronym{C-V2X}{C-V2X}{Cellular-V2X}     
\newacronym{NR}{NR}{New Radio}              
\newacronym{ETSI}{ETSI}{European Telecommunications Standards Institute}  
\newacronym{EHF}{EHF}{extremely high-frequency}
\newacronym{BER}{BER}{bit-error-rate}
\newacronym{ICI}{ICI}{inter-carrier interference}
\newacronym{SotA}{SotA}{state-of-the-art}
\newacronym{DoF}{DoF}{degrees-of-freedom}
\newacronym{UAV}{UAV}{unmanned aerial vehicle}
\newacronym{CAV}{CAV}{connected autonomous vehicles}
\newacronym{MIMO}{MIMO}{multiple-input multiple-output}
\newacronym{SISO}{SISO}{single-input single-output}
\newacronym{UE}{UE}{user equipment}
\newacronym{AP}{AP}{access point}
\newacronym{CIR}{CIR}{channel impulse response}
\newacronym{CSI}{CSI}{channel state information}
\newacronym{SNR}{SNR}{signal-to-noise ratio}
\newacronym{AWGN}{AWGN}{additive white Gaussian noise}
\newacronym{ML}{ML}{maximum likelihood}
\newacronym{OOB}{OOB}{out-of-band}
\newacronym{MX}{MX}{multiplexing}
\newacronym{DMX}{DMX}{demultiplexing}
\newacronym{AoA}{AoA}{angle-of-arrival}
\newacronym{AoD}{AoD}{angle-of-departure}
\newacronym{5GAA}{5GAA}{5G Automotive Association} 
\newacronym{6G-IA}{6G-IA}{6G Smart Networks and Services Industry Association}  
\newacronym{MUSIC}{MUSIC}{MUltiple SIgnal Classification}
\newacronym{ESPRIT}{ESPRIT}{estimation of signal parameters via rotational invariance techniques}
\newacronym{DP}{DP}{detection problem}
\newacronym{EP}{EP}{estimation problem}
\newacronym{KPI}{KPI}{key performance indicator}
\newacronym{ISI}{ISI}{inter-symbol interference}
\newacronym{TVIRF}{TVIRF}{time-variant impulse response function}
\newacronym{TVTF}{TVTF}{time-variant transfer function}
\newacronym{DVIRF}{DVIRF}{Doppler-variant impulse response function}
\newacronym{DVTF}{DVTF}{Doppler-variant transfer function}
\newacronym{DDSF}{DDSF}{delay-Doppler spread function}
\newacronym{PAPR}{PAPR}{peak-to-average power ratio}
\newacronym{RCS}{RCS}{radar cross-section}
\newacronym{PIR}{PIR}{peak interference residual}
\newacronym{PSR}{PSR}{peak-to-sidelobe ratio}
\newacronym{PA}{PA}{power amplifier}
\newacronym{RF}{RF}{radio frequency}
\newacronym{CNIT}{CNIT}{Consorzio Nazionale Interuniversitario per le Telecomunicazioni}
\newacronym{PNRR}{PNRR}{Italian National Recovery and Resilience Plan}
\newacronym{DT}{DT}{discrete time}
\newacronym{CT}{CT}{continuous time}
\newacronym{UAVs}{UAVs}{unmanned aerial vehicles}
\newacronym{TF}{TF}{severe time-frequency}
\newacronym{DD}{DD}{delay-Doppler}
\newacronym{AFD}{AFD}{affine Fourier domain}
\newacronym{PHY}{PHY}{physical-layer}
\newacronym{FTN}{FTN}{faster-than-Nyquist}
\newacronym{SEFDM}{SEFDM}{spectrally efficient frequency division multiplexing}
\newacronym{HWIs}{HWIs}{hardware impairments}
\newacronym{PN}{PN}{phase noise}
\newacronym{PMEPR}{PMEPR}{peak-to-mean envelope power ratio}
\newacronym{RRC}{RRC}{root-raised cosine}
\newacronym{DCP}{DCP}{double CP}
\newacronym{TBI}{TBI}{time-burst interference}
\newacronym{NBI}{NBI}{narrowband interference}
\newacronym{LO}{LO}{local oscillator}
\newacronym{CFO}{CFO}{carrier frequency offset}
\newacronym{AFS}{AFS}{affine Fourier series}
\newacronym{PSD}{PSD}{power spectral density}
\newacronym{SJ}{SJ}{sampling jitter}
\newacronym{CRBs}{CRBs}{Cram{\'e}r-Rao Bounds}
\newacronym{TX}{TX}{transmit}
\newacronym{RX}{RX}{receive}
\newacronym{TD}{TD}{time domain}
\newacronym{QAM}{QAM}{quadrature amplitude modulation}
\newacronym{RHS}{RHS}{right-hand side}
\newacronym{CFT}{CFT}{continuous-time Fourier transform}
\newacronym{SI}{SI}{self-interference}
\newacronym{IR}{IR}{impulse response}
\newacronym{LMMSE}{LMMSE}{linear minimum mean square error}
\newacronym{PAM}{PAM}{pulse amplitude modulation}
\newacronym{FD}{FD}{frequency domain}
\newacronym{SINR}{SINR}{signal-to-interference-plus-noise ratio}
\newacronym{PSK}{PSK}{phase-shift keying}
\newacronym{FIM}{FIM}{Fisher information matrix}
\newacronym{PDP}{PDP}{power delay profile}
\newacronym{FS}{FS}{Fourier series}
\newacronym{HWI}{HWI}{hardware impairment}
\newacronym{CRB}{CRB}{Cram\'{e}r-Rao Bound}
\newacronym{std}{std}{standard deviation}
\newacronym{CPP}{CPP}{chirp-periodic prefix}
\newacronym{SC}{SC}{suppressed carrier}